\PassOptionsToPackage{hyphens}{url}
\documentclass[11pt]{article}

\usepackage[preprint]{acl}

\usepackage{times}
\usepackage{latexsym}

\usepackage[T1]{fontenc}

\usepackage[utf8]{inputenc}

\usepackage{microtype}

\usepackage{inconsolata}

\usepackage{graphicx}

\usepackage{booktabs}       
\usepackage{amsfonts}       
\usepackage{nicefrac}       
\usepackage{microtype}      
\usepackage{xcolor}         
\usepackage{amsmath,amssymb,amsfonts,amsthm}
\usepackage[tikz,xcolor]{mdframed}
\surroundwithmdframed[roundcorner=5pt]{problem}

\usepackage[capitalize]{cleveref}
\crefname{table}{Tab.}{Tabs.}
\Crefname{problem}{Problem}{Problems}
\Crefname{finding}{Finding}{Findings}
\Crefname{table}{Table}{Tables}
\crefname{figure}{Fig.}{Figs.}
\Crefname{figure}{Figure}{Figures}
\crefname{section}{Sec.}{Secs.}
\Crefname{section}{Section}{Sections}
\crefname{algorithm}{Alg.}{Algs.}
\Crefname{algorithm}{Algorithm}{Algorithms}

\usepackage{bm}

\usepackage{booktabs}
\usepackage{multirow}
\usepackage{makecell}
\usepackage{threeparttable}

\usepackage{graphicx}
\usepackage{subcaption}

\usepackage{tabularx}
\newcolumntype{Y}{>{\centering\arraybackslash}X}

\usepackage[dvipsnames]{xcolor}

\usepackage{colortbl}
\definecolor{LightGray}{gray}{0.8}

\usepackage{calc}

\usepackage{enumitem}

\usepackage{lipsum}

\usepackage{dirtree}

\usepackage[hang,flushmargin]{footmisc}

\usepackage{listings}
\lstdefinestyle{promptstyle}{
    basicstyle=\tiny\ttfamily,
    breaklines=true,
    frame=single,
    backgroundcolor=\color{gray!10},
    numberstyle=\tiny,
    linewidth=\linewidth,
}

\usepackage{algorithmic}
\usepackage[ruled,linesnumbered]{algorithm2e}
\SetAlCapFnt{\small}
\SetAlCapNameFnt{\small}

\title{Defending FCG-Based Malware Detectors Against Metamorphic Android Variants via LLM-Powered Code Refactoring}

\author{
 \textbf{Ngoc N. Tran\textsuperscript{1}},
 \textbf{Noah Dahle\textsuperscript{1}},
 \textbf{Bo Ni\textsuperscript{1}},
 \textbf{Anwar Said\textsuperscript{2}},
 \\
 \textbf{Waseem Abbas\textsuperscript{3}},
 \textbf{Xenofon D. Koutsoukos\textsuperscript{1}},
 \textbf{Tyler Derr\textsuperscript{1}}
\\
\\
 \textsuperscript{1}Vanderbilt University
    \;
 \textsuperscript{2}Highmark Health
    \;
 \textsuperscript{3}University of Texas at Dallas
}

\begin{document}
\maketitle

\begin{abstract}
LLM-generated malwares pose a mounting threat to automated detection systems, as rewritten malicious code shifts beyond what classifiers learned from static corpora.
While graph-based Android malware classifiers operating on function call graphs, especially when enriched with LLM features, have established strong detection accuracy; they remain brittle against this emerging attack surface, dropping up to over 60\% in accuracy.
Standard graph augmentation strategies fail to address this gap: they are context-agnostic, producing graphs that are incoherent as programs and bear no resemblance to these LLM-rewrite attacks.
To counter this threat, we propose leveraging LLMs to synthesize code augmentations for diverse, realistic training variants.
We introduce FCGA, a framework operating at two complementary levels: rewriting individual functions into equivalent yet distinct implementations, and jointly refactoring clusters of functions via summarization-guided prompting, concentrating on the most important graph components. 
Across a controlled real-world OOD benchmark and a supporting direct LLM-rewrite stress test, FCGA enhances robustness by up to 10\% over augmentation baselines. Our analysis further reveals a fidelity-diversity tradeoff that determines which augmentation strategies are most effective during pretraining and adaptation.
The source code is available at \url{https://github.com/ngoctnq/fcga}.
\end{abstract}

\section{Introduction}

Android malware is a persistent and escalating security threat, with millions of malicious applications discovered each year~\cite{blogKeepingGoogle}.
The problem has been exacerbated by the rapid advancement of large language models (LLM): attackers can now prompt LLMs to rewrite malicious code into functionally equivalent yet structurally distinct implementations at scale, generating diverse evasive variants that are difficult for automated detectors to catch~\cite{akil2025llmalmorphfeasibilitygeneratingvariant,hortea2026infinitemutationenginemeasuring}.
In response, significant research effort has gone into building robust malware detectors, with graph classifiers operating on Function Call Graphs (FCGs) have emerged as the leading approach~\cite{freitas2021malnet,liu2024segdroid} as they faithfully models how malwares operate while remaining tractable even with large real-world applications~\cite{mehadi2017scalable}.

Recent work has further shown that enriching FCG nodes with LLM-generated code embeddings, which inject semantic information into this structural-only representation, improves both detection accuracy and generalization to unseen malwares~\cite{mntf}.
Yet even with these advances, we discover that classifiers still struggle against LLM-rewritten variants not represented during training, suffering significant performance degradation of over 60\% in accuracy in some cases.

A central practical challenge is limited labeled coverage of emerging malware variants, i.e., a major cause of brittleness is distributional, stemming from a lack of both data and diversity~\cite{manzanares2022concept}. Classifiers rely on fixed, manually annotated corpora that cannot capture rapidly evolving implementations.
When an attacker uses an LLM to rewrite a malware sample, the resulting representation can shift sufficiently far from the training distribution to evade detection, even though the underlying malicious behavior is preserved.
We therefore first consider standard, problem-agnostic graph augmentations to address data scarcity; however, their semantic-agnostic operations do not reflect the implementation changes produced by LLM-based attackers.
These augmented graphs occupy a distribution entirely different from LLM-rewritten malwares, which are semantically-shifted from the training distribution, and fail to prepare for this threat practitioners face.

\begin{figure}[t]
    \centering
    \includegraphics[width=0.8\linewidth]{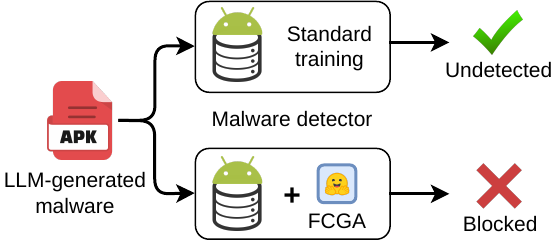}
    \caption{Illustration of the problem.
    FCGA leverages LLMs to generate semantically-grounded augmentations closer to the distribution of LLM-based variants, enhancing resilience against this threat.}
    \label{fig:problem}
    \vspace{-0.5em}
\end{figure}

To address this problem, we seek to investigate the extent of our ability to defend against LLM-generated malware by proactively utilizing the same tool that attackers use for a principled response to this attack surface.
Our core thesis is that the same LLM-driven refactoring capability used to create evasive variants can be repurposed to strengthen the defender's training distribution.
Concretely, we explore generating LLM variants as training augmentations, resolving data scarcity at its source by directly expanding the training corpus to cover LLM-rewritten distributions. Given the effectiveness of LLM-augmented FCG classification, we further leverage the power of LLMs to generate new meaningful labeled examples without the need for human annotation.
\Cref{fig:problem} illustrates the problem, and how our proposed framework resolves~it.

Our contributions are as follows:
\begin{itemize}
    \item We introduce a proactive defense paradigm that repurposes LLM-driven code refactoring to prepare malware classifiers for emerging and previously unseen variants.

    \item We propose FCGA, a code-refactoring-guided training augmentation framework operating at both the function and FCG-subgraph levels to generate semantically and structurally diverse training variants.

    \item We develop structural and model-aware selection strategies that concentrate augmentation on graph components most influential to program structure or the target classifier.

    \item Across four GNN architectures and multiple OOD settings, the best FCGA variants improve robustness over unaugmented and generic graph-augmentation baselines, including an 10\% gain on unseen LLM-rewrite attack. Our analysis further identifies a fidelity-diversity tradeoff: broader structural variation is particularly useful for robust pretraining, while targeted, model-aware augmentation becomes more valuable during adaptation.
\end{itemize}

\section{Preliminaries and Related Works}

\subsection{Graph Representation for Code Analysis}

Graphs are a natural representation for code: programs are inherently structured by syntactic nesting, control flow, and inter-procedural dependencies, all of which can be directly encoded as nodes and edges~\cite{allamanis2018survey}.
Traditional program analysis tasks extract code graphs at the function level~\cite{alon2019code2vec,zhou2019devign} for intra-procedural reasoning, and then unifying these local structures with call edges for a holistic view of the entire program~\cite{yamaguchi2014modeling}.
While expressive, they do not scale to real-world applications, where a single program can produce graphs with millions of nodes that are intractable for GNN inference.
FCGs have emerged as the standard graph representation for binary analysis at scale~\cite{freitas2021malnet} by taking a coarser inter-procedural view, where each node represents one function, and each directed edge indicates an invocation from the caller to the callee.
As a tradeoff, they omit details about individual functions: each node is reduced to an opaque identifier with no signal about its underlying functionality; motivating the need for richer node embeddings to offset this loss of information~\cite{mntf}.

\subsection{Graph Data Augmentation}
\label{sec:graph-aug}

Data augmentation for graphs aims to diversify the training distribution by generating synthetic graph variants, improving generalization without additional labeled data~\cite{zhao2022graph}.
Existing methods broadly fall into two categories: structural augmentations, which perturb the graph topology by randomly dropping nodes~\cite{you2020graphcl}, edges~\cite{rong2020dropedge}, or sampling connected subgraphs as views~\cite{you2020graphcl}; and feature augmentations, which instead perturb node or edge attributes, for example by randomly masking feature dimensions or injecting noise into embedding vectors~\cite{you2020graphcl}.

For code graphs, however, these approaches fall short.
Random augmentations operate without any understanding of what nodes and edges represent, and thus would produce graphs that are structurally implausible and semantically incoherent as programs.
This limits the diversity and realism of the augmented training set, motivating our augmentation strategy that leverages functional semantics via LLM code comprehension.

\subsection{LLMs for Code Rewrites}
At a high level, LLMs can be prompted to produce semantically-equivalent reimplementations from natural-language instructions.
In the security literature, LLM-driven code rewriting has been explored primarily as an attack surface.
\citet{Madani_2023} demonstrated that LLMs can serve as embedded mutation engines for evasive metamorphic malware.
\citet{akil2025llmalmorphfeasibilitygeneratingvariant} showed that function-level LLM rewrites can achieve up to 91\% attack success rate against an ML-based malware \mbox{detector.}
In software engineering, LLM rewrites are used to improve models rather than to evade them.
\citet{ouchebara2026llmbasedvulnerablecodeaugmentation} leveraged LLMs to augment vulnerable code datasets, through both generation and refactoring, to train more robust vulnerability detectors.
\citet{wang2023recode} evaluated the robustness of code generation models against a taxonomy of semantics-preserving transformations.
This dual-use is central to our approach: the same generative capability that attackers exploit to produce evasive variants can be harnessed by defenders to proactively synthesize diverse training data.
\section{Problem Formulation}

We formulate the problem of robust Android malware classification as a graph-level classification task.
Each Android application is represented as a directed graph $\mathcal{G} = (\bm{X}, \bm{A})$, where $\bm{X}\in\mathbb{R}^{|V|\times d}$ are the extracted features of the functions in said application, and $\bm{A}\in\mathbb{R}^{|V|\times |V|}$ is the adjacency matrix of the FCG
encoding caller-callee relationships, $V$ being the set of function nodes, and $d$ is the dimension of the source code embeddings.
The task is graph classification: given a labeled training set $\mathcal{D} = \{(\mathcal{G}_i, y_i)\}_{i=1}^N$, train a GNN classifier $f_\theta$ to predict the true malware family label $y \in \mathcal{Y}$ for a graph $\mathcal{G}$, robust to the attacker's evasion attempt $\mathcal{G}'$ towards an incorrect prediction $y'\ne y$.
\section{Methodology}
\label{sec:method}
In this section, we describe the desired properties for a solution to the problem (\Cref{sec:motivation}), and how our design choices address them across the two granularities that FCGA operates on: function-level (\Cref{sec:function-level}), and subgraph-level (\Cref{sec:subgraph-level}).

\begin{figure*}
    \centering
    \includegraphics[width=\linewidth]{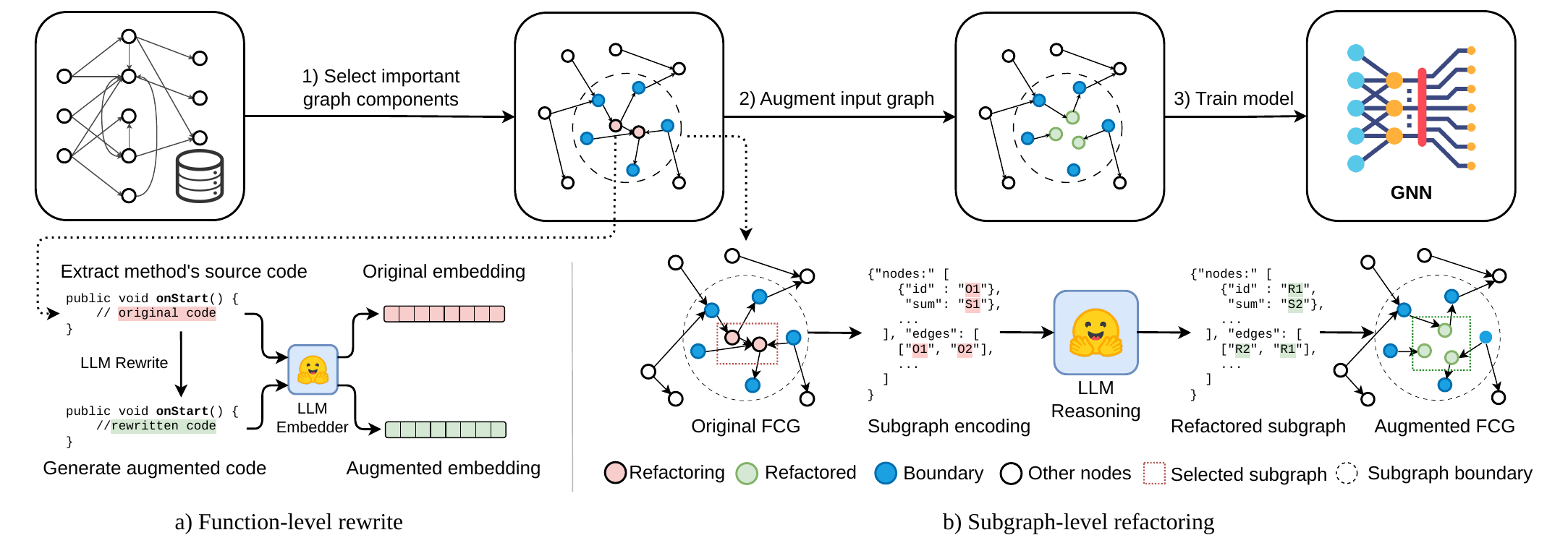}
    \vspace{-2em}
    \caption{Diagram of our augmentation pipeline. For each training sample, FCGA first selects important graph components to be augmented based on the original sample and the target model architecture. These components are then modified to produce a new augmented training sample, which is used as additional data during model training.}
    \label{fig:pipeline}
    \vskip -1.5ex
\end{figure*}

\subsection{Motivation}
\label{sec:motivation}
We identify three axes along which a desired augmentation approach must operate:

\paragraph{Implementation variation.}

Varying how a function is implemented prevents $f_\theta$ from overfitting to specific coding conventions that may not generalize across implementations of the same behavior.
While standard perturbation-based augmentations can introduce noise to $\bm{X}$, they lack semantic grounding and thus may not produce meaningful variation in the sample space:
because there is no well-defined inverse mapping from a perturbed embedding back to source code, the shifted node features may not correspond to any plausible alternative implementation.
In contrast, LLM-driven rewriting is explicitly prompted to preserve functionality while producing structurally-distinct, semantically-informed training variants, resulting in samples that are both realistic and diverse.

\paragraph{Structural diversity.}

By changing the program's call structure and module boundaries, we can proactively improve the robustness of our model to refactoring-based evasion, where an attacker restructures the program's functional decomposition without changing its high-level behavior~\cite{choliy2017obfuscating,yong2024understanding}.
Augmentation along this axis requires modifying $\bm{A}$ directly, producing variants whose FCG topology differs from the original. Existing structural augmentations can introduce such variation, but besides sharing the same improbable generation weakness as feature perturbation methods, they could remove informative components of the original graph that are crucial for the model's performance.

\paragraph{Model awareness.}

Traditionally, these important nodes are identified by analyzing the graph structure alone, e.g. by selecting high-degree nodes or central hubs.
However, the GNN may rely on a different subset of nodes that are not necessarily the most structurally prominent: a low-degree function that implements a critical piece of logic may be more influential to the model's prediction than a high-degree utility function.
By leveraging the trained $f_\theta$'s own signals to identify which nodes it relies on, we can ensure that augmentation efforts are focused on the components that matter most to the model's decision boundary.

\subsection{Function Call Graph Augmentation}
Following these insights, we propose our augmentation strategy with two different variations.
Each is composed of two stages: first, selecting important graph components to be augmented:
given the scale of real-world FCGs, which can contain thousands of nodes~\cite{freitas2021malnet}, augmenting every complete graph is computationally intractable and unnecessary. Afterwards, the selected components are modified to produce a new augmented training sample, which is used as additional data during model training.
\Cref{fig:pipeline} illustrates the overall pipeline of our method, and \Cref{tab:baselines} summarizes how our variations compare against augmentation baselines with respect to the three identified axes.

\newcommand{\greyno}{\textcolor{LightGray}{-}}
\begin{table}[t]
\footnotesize 
\vspace{1.5ex}
\centering
\begin{tabularx}{\linewidth}{Xccc}
\toprule
Method & Structure & Semantics  & Model-aware \\
\midrule
Baseline & \greyno & \greyno    & \greyno \\
\midrule
NodeDrop & \checkmark & \greyno    & \greyno \\
EdgeDrop & \checkmark & \greyno    & \greyno \\
Node+Edge & \checkmark & \greyno    & \greyno \\
\midrule
Function & \greyno & \checkmark & \greyno \\
\mbox{... + Model} & \greyno & \checkmark & \checkmark \\
\midrule
Perturb & \greyno & \textasciitilde & \greyno \\
\mbox{... + Model} & \greyno & \textasciitilde & \checkmark \\
\mbox{... + All} & \greyno & \textasciitilde & \greyno \\
\midrule
Subgraph & \checkmark & \checkmark    & \greyno \\
\mbox{... + Model} & \checkmark & \checkmark    & \checkmark \\
\bottomrule
\end{tabularx}
\caption{Summary of augmentation methods, including baselines defined in Section~\ref{sec:graph-aug} and our FCGA variants, with their utilized information.}
\label{tab:baselines}
\end{table}

\subsubsection{Function-Level Code Rewriting}
\label{sec:function-level}

Function-level FCGA (FCGA-F) augments the FCG at the node level: for each selected function, the LLM rewrites its decompiled source code under explicit semantic-preservation instructions, producing an alternative implementation whose embedding replaces the original node feature.
The graph topology remains unchanged; only feature representations of selected nodes are modified. 
The overall FCGA-F pipeline is illustrated in \Cref{fig:pipeline}a.

\paragraph{Node selection.}
Our node selection criteria are divided into two families: static topological measures that are derived from the input graph structure, and model-dependent measures that leverage the trained GNN's own signals. Each option has its own advantages: the former can be efficiently pre-computed offline without any reference model, while the latter directly identifies nodes that are influential to the model's decision boundary. 
Additional criteria details are in Appendix~\ref{app:selection-criteria}.

\textit{1) Structure-based.}
Static centrality measures can be used to capture structural prominence in the FCG without requiring a trained model, where each of them measures a different sense of node importance to the overall network positioning.
These high-centrality nodes greatly contribute to the critical information flows during message passing, and thus are prime candidates for augmentation.

\textit{2) Model-based.}
While these degree-based measures can detect nodes that are critical to the general information flow through the graph, they need not coincide with the actual propagation determined by the GNN's weights.
Instead of statically relying on the input graph structure, we can leverage the model $f_\theta$ to identify nodes that are most influential to its predictions, making them more directly relevant to augmenting the model's decision boundary. We thus borrow prominent methods from the explainable graph learning literature~\cite{ying2019gnnexplainergeneratingexplanationsgraph,vu2020pgmexplainerprobabilisticgraphicalmodel} to accomplish this task.

\vspace{-1ex}
\paragraph{Augmentation.}
From the selected nodes, we can start generating augmented samples by modifying their code and thus their embeddings, while keeping the rest of the graph intact. We discuss our specific FCGA-F prompting strategy, and then introduce a cheaper approximating alternative.

\textit{1) Function-rewrite.}
The prompting strategy balances preservation with transformation.
We impose two invariants: (1) the function signature must remain intact, ensuring that the FCG's caller-callee edges are unaffected; and (2) the function must exhibit the same observable behavior, i.e. \mbox{identical} outputs and side effects for all inputs.
Within these constraints, the LLM is encouraged to apply maximum syntactic transformation: restructuring control flow, renaming variables, reorganizing computations, and substituting equivalent Java API calls.
This produces rewrites that occupy a diverse region of the embedding space while preserving the function's semantic identity.
These constraints guide generation rather than provide formal equivalence guarantees; the rewritten code is used only to produce alternative node representations for training.

\textit{2) Embedding perturbation.}
From the insight that similar code yields similar embeddings, we can also cheaply generate augmented samples by directly perturbing node features in the embedding space.
Perturbation-based FCGA (FCGA-P) augments node features by adding a Gaussian noise vector to the original embedding of each selected node, scaled analytically so that the perturbed vector lies at a prescribed cosine similarity from the original.
This makes FCGA-P substantially cheaper to apply as it requires no LLM calls for rewrite nor re-embedding, and can be computed in closed-form on-the-fly during training.
We can further leverage FCGA-P to augment the full training set, generating a completely new set of augmented samples at every training iteration for a more powerful pre-training effect, reserving the more costly FCGA-F for fine-tuning when the model is more sensitive to semantically meaningful variation.

\subsubsection{Subgraph-Level Code Refactoring}
\label{sec:subgraph-level}
In contrast to FCGA-F/P, FCGA-S jointly refactors a cluster of functions, reshaping the FCG topology itself via splitting, merging, or reorganizing functions within the subgraph, while being prompted to preserve the intended high-level behavior and malware-family identity of the represented sub-program.
By not strictly replicating the code graph, this approach covers a broader space of program organizations than its function-level counterparts.
The FCGA-S pipeline is illustrated in \Cref{fig:pipeline}b.

\paragraph{Node selection.}
A downside of function-level selection is that they can only measure individual graph components in isolation, and thus fail to capture the joint influence of multiple nodes that may be individually unimportant but collectively important.
Moreover, tightly connected clusters are preferred candidates because they tend to represent self-contained, cohesive functionality; making them more amenable to successful LLM refactoring.
We thus utilize subgraph selection to form a localized neighborhood for joint refactoring.

\textit{1) Structure-based.}
We can adapt any structure-based graph component importance measures by using Monte Carlo Tree Search (MCTS).
From an empty set, we gradually build the subgraph by adding adjacent nodes, enforcing connectivity of the selected candidates.
The candidate subgraph reward value can be computed by collapsing itself into a single node, then re-computing the resulting new graph's node centrality metrics.

\textit{2) Model-based.}
Instead of previous explainable AI methods identifying nodes independently, we opt for SubgraphX~\cite{yuan2021explainability}, which finds important subgraphs utilizing MCTS with Shapley values~\cite{lundberg2017unified} as the reward signal, which attribute the model's output to each subgraph by computing its weighted average marginal contribution across all possible subsets.

\paragraph{Augmentation.}
FCGA-S operates in three stages.
First, for each node in the selected subgraph and upto one-hop away, the LLM generates a concise summary of its behavior.
These summaries shift the LLM's reasoning from \textit{how} a function is written to \textit{what} it does, enabling more principled structural decisions, while compressing the input to fit within the LLM's context window.
Second, the subgraph is presented to the LLM
by its node signatures, summaries, edges, and one-hop context; and the LLM is given full creative freedom to split, merge, rename, or reorganize the selected nodes while keeping boundary context nodes intact, ensuring the resulting subgraph is mendable back into the original FCG without disrupting the wider program logic.
Finally, the augmented data sample is reconstructed by replacing the original subgraph with the refactored subgraph, which summary embeddings as node features.
Given the embedder's capability to capture semantic similarity of source codes and their descriptions as it was demonstrated in code retrieval tasks~\cite{qwen3embedding}, the new summaries' embedding distribution is expected to be close to the intended implementation, while the structural changes introduce additional diversity that is orthogonal to function-level rewriting.
\begin{table*}[t]
\small
\centering
\begin{tabularx}{\textwidth}{lYYYYYYYY}
\toprule
\multirow{2.5}{*}{Method} & \multicolumn{2}{c}{GCN} & \multicolumn{2}{c}{GIN} & \multicolumn{2}{c}{GAT} & \multicolumn{2}{c}{GraphSAGE} \\
\cmidrule(lr){2-3} \cmidrule(lr){4-5} \cmidrule(lr){6-7} \cmidrule(lr){8-9}
& Test & OOD & Test & OOD & Test & OOD & Test & OOD \\
\midrule
Baseline &
    $94.57_{0.19}$ & $54.93_{1.20}$ &
    $94.70_{0.08}$ & $57.77_{1.60}$ &
    $92.13_{0.74}$ & $56.53_{1.31}$ &
    $94.37_{0.66}$ & $58.57_{0.54}$ \\
\midrule
NodeDrop &
    $94.67_{0.68}$ & $55.90_{2.86}$ &
	$94.67_{0.05}$ & $58.53_{1.58}$ &
	$91.20_{2.09}$ & $52.63_{4.71}$ &
    $94.40_{0.46}$ & $58.60_{1.57}$ \\
EdgeDrop &
	$94.43_{0.08}$ & $55.57_{1.80}$ &
	$94.87_{0.50}$ & $58.97_{3.52}$ &
	$89.53_{3.68}$ & $54.77_{4.45}$ &
    $94.37_{0.50}$ & $57.77_{3.02}$ \\
Node+Edge &
	$94.50_{0.36}$ & $56.33_{0.71}$ &
	$94.50_{0.26}$ & $58.87_{2.59}$ &
	$90.93_{2.19}$ & $57.57_{2.66}$ &
    $94.53_{0.14}$ & $58.63_{1.45}$ \\
\midrule
Function &
	$94.73_{0.67}$ & $56.27_{3.25}$ &
    $94.97_{0.40}$ & \underline{$60.97_{3.66}$} &
    $92.20_{1.41}$ & $58.63_{2.85}$ &
    $94.67_{0.40}$ & $58.77_{0.70}$ \\
... + Model &
	$94.67_{0.12}$ & $55.97_{3.09}$ &
	$94.90_{0.14}$ & $59.73_{1.27}$ &
	$93.03_{0.21}$ & $59.57_{0.86}$ &
    $94.50_{0.80}$ & $58.93_{0.59}$ \\
\midrule
Perturb &
	$94.40_{0.09}$ & $56.13_{1.36}$ &
	$94.90_{0.26}$ & $60.43_{1.40}$ &
	$92.43_{0.38}$ & $58.80_{4.00}$ &
    $94.47_{0.31}$ & \underline{$60.37_{5.10}$} \\
... + Model &
	$94.57_{0.79}$ & $\mathbf{57.73}_\mathbf{3.09}$ &
	$\mathbf{95.13}_{0.42}$ & $60.30_{0.90}$ &
	$\mathbf{93.80}_\mathbf{0.17}$ & $\mathbf{60.53}_\mathbf{0.76}$ &
    $94.40_{0.21}$ & $58.77_{0.84}$ \\
... + All &
	$\mathbf{94.90}_\mathbf{0.43}$ & $56.47_{1.86}$ &
	$\mathbf{95.13}_\mathbf{0.24}$ & $60.90_{2.63}$ &
	$92.33_{4.29}$ & $59.60_{1.35}$ &
    $\mathbf{94.77}_\mathbf{0.21}$ & $58.53_{0.33}$ \\
\midrule
Subgraph &
	$94.37_{0.16}$ & \underline{$56.67_{2.16}$} &
	$94.33_{0.05}$ & $60.33_{2.50}$ &
	$92.27_{3.64}$ & $58.37_{4.23}$ &
    $94.73_{0.24}$ & $59.27_{0.45}$ \\
... + Model &
	$94.77_{0.24}$ & $56.17_{2.75}$ &
	$94.93_{0.21}$ & $\mathbf{62.47}_\mathbf{2.82}$ &
    $91.90_{0.12}$ & \underline{$60.30_{3.49}$} &
    $93.97_{0.19}$ & $\mathbf{60.73}_\mathbf{0.45}$ \\
\bottomrule
\end{tabularx}
\caption{Models' test set accuracy in percentage on the original and covariance-shifted MalNet-Tiny datasets across different augmentation strategies (higher is better).
Results are averaged over 3 seeds, subscripts denote standard deviations.
Largest value in each column in bold, second largest OOD underlined.
}
\label{tab:pretrain}
\end{table*}

\section{Experiments}
\label{sec:experiments}

In this section, we evaluate the effectiveness of different augmentation strategies on real-world Android malware datasets.
We describe the experimental setup, including datasets, evaluation protocols, and baselines, then present our results.

\subsection{Experimental Setup}
For a comprehensive description, please refer to Appendix~\ref{app:implementation} for the designs and implementation, and Appendix~\ref{app:ablations} for hyperparameter selection.

\subsubsection{Datasets}
We evaluate on MNTF~\cite{mntf}, an OOD benchmark for Android malwares, specifically its \textit{Tiny} reference split, and \textit{Common} split modeling for covariate shift.
Each application is represented as an FCG whose nodes carry 1024-dimensional code embeddings produced by Qwen3-Embedding-0.6B~\cite{qwen3technicalreport} from decompiled Java source, and labeled by its malware family.
We consider two evaluation protocols:
\textit{Test} performance is measured on the \textit{Tiny} test split to measure generalization to unseen data of the same distribution, while \textit{OOD} generalization is measured on the \textit{Common} split, evaluating the model's ability to new variants.

\begin{figure*}[t]
    \centering
    \vskip 0.5ex
    \begin{subfigure}[t]{0.50\linewidth}
        \centering
        \includegraphics[width=\linewidth]{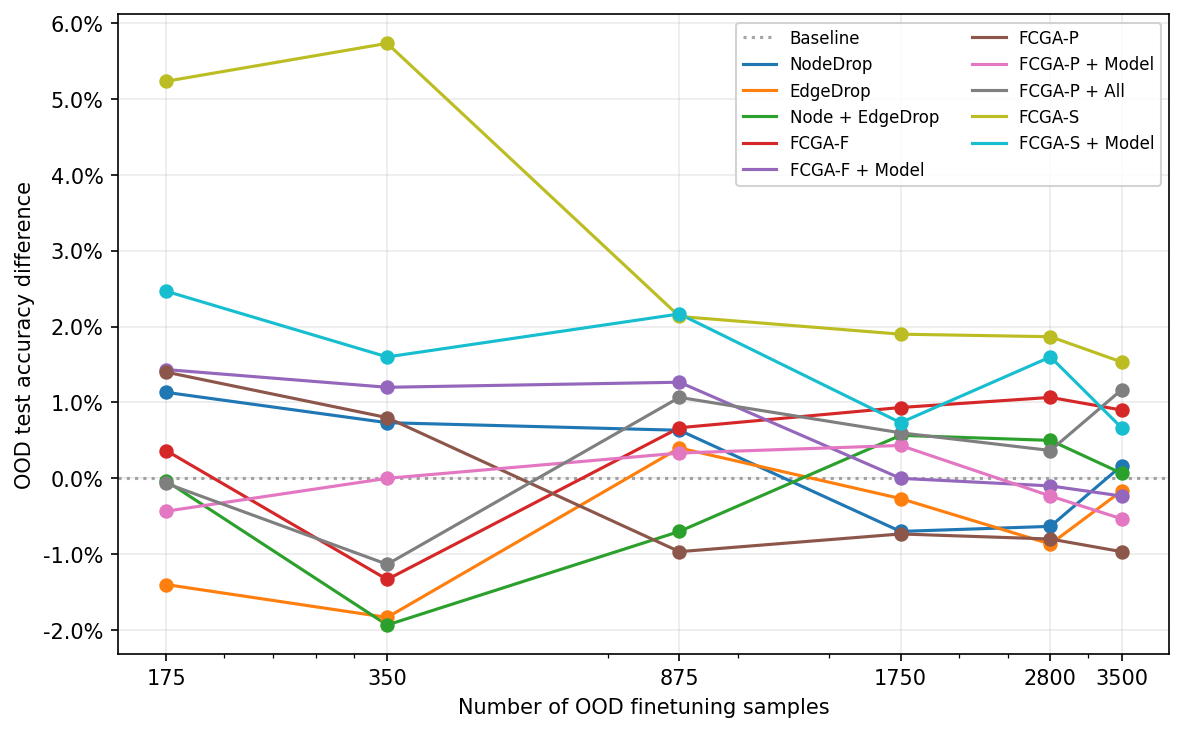}
        \vskip 0.5ex
        \caption{Augmented Pre-Training + Standard Fine-Tuning}
        \label{fig:aug-base}
    \end{subfigure}%
    ~ 
    \begin{subfigure}[t]{0.50\linewidth}
        \centering
        \includegraphics[width=\linewidth]{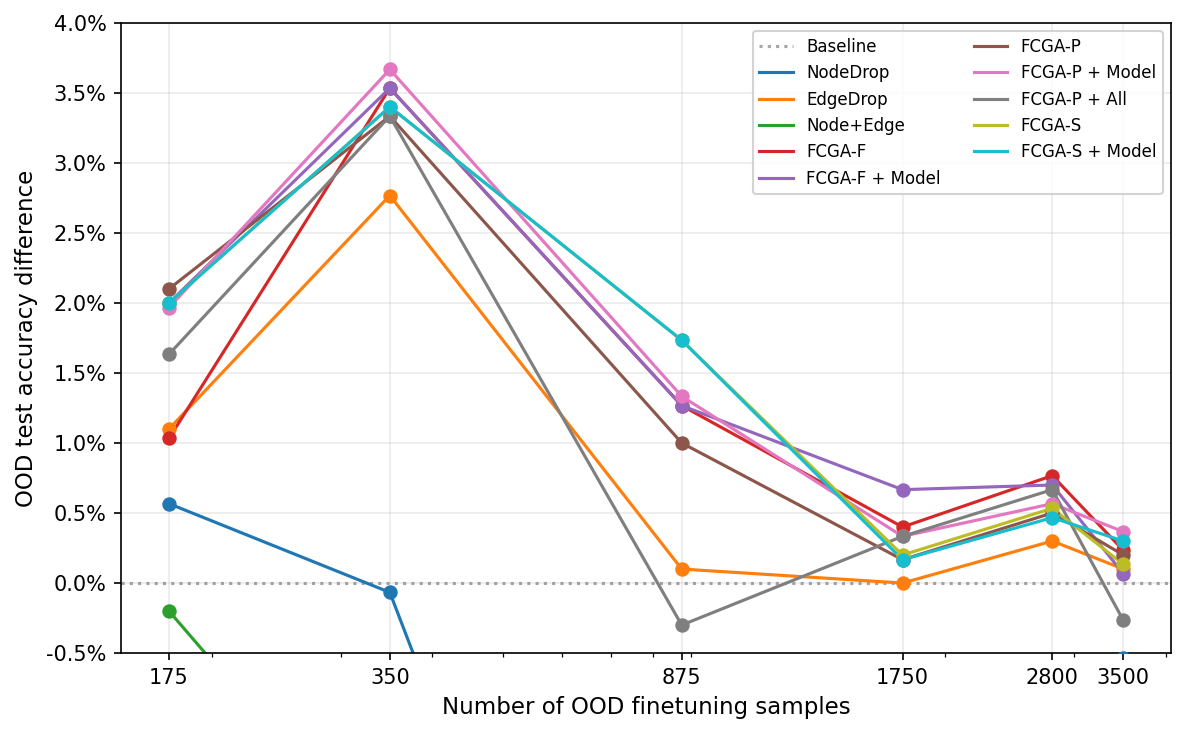}
                \vskip 0.5ex
        \caption{Standard Pre-Training + Augmented Fine-Tuning}
        \label{fig:aug-ft}
    \end{subfigure}
    \caption{OOD accuracy difference versus unaugmented baseline after finetuning on a pretrained GIN model, ablated across different augmentation strategies.
    Higher is better.}
    \label{fig:ft-truncate}
    \vskip 0.5ex
\end{figure*}

\subsubsection{Architectures \& Baselines}
We evaluate across multiple GNN architectures to ensure findings generalize beyond a single backbone: GCN~\cite{kipf2017semisupervisedclassificationgraphconvolutional}, GIN~\cite{xu2019powerfulgraphneuralnetworks}, GAT~\cite{veličković2018graph}, and GraphSAGE~\cite{hamilton2018inductiverepresentationlearninglarge}.

We compare FCGA against the following augmentation strategies, spanning structure-only and semantic-driven approaches.
\textbf{Baseline} denotes the unaugmented performance without any augmentation.
\textbf{NodeDrop}~\cite{you2020graphcl} and \textbf{EdgeDrop}~\cite{rong2020dropedge} randomly remove nodes and edges, respectively; and \textbf{Node+Edge} applies both simultaneously.

\subsection{Results \& Evaluation}

We consider two scenarios: off-the-shelf pre-training, where the resulting model is directly used for evaluation; and downstream fine-tuning, where the existing checkpoint is then fitted to a limited set of new labeled data.
Additional settings are studied in Appendix~\ref{app:ablations} and \ref{app:discussions} in the interest of space.

\subsubsection{Augmented Pre-Training}
\label{sec:pretraining}

\Cref{tab:pretrain} shows the results of augmented pre-training across different augmentation strategies and GNN architectures. We observe that FCGA variants consistently outperform all baselines across architectures and evaluations, demonstrating the effectiveness of our context-aware augmentation strategy in both in-domain and OOD generalization.

For in-domain test performance, accuracies saturated across all models.
Regardless, FCGA-P still achieves the highest test accuracy across all architectures, showing that even in this setting with little possible gain, our method can still provide up to 1.7\% accuracy increase over the unaugmented baseline. This supports our claim that FCGA-P can be used as a cheap general-purpose augmentation strategy, having its semantic-aware edge over traditional context-agnostic approaches.

In the OOD setting where models struggle to generalize, we find that model-aware augmentation does not provide significant benefits for FCGA-F, but much more effective for FCGA-P/S. This implies that a stronger augmentation strategy is required when targeting the malware-carrying nodes: we discuss this tradeoff in more detail in \Cref{sec:tradeoffs}. 
Meanwhile, model-aware variants
strictly yield the best OOD performance across all architectures, even higher than FCGA-P + All, highlighting the power of altering higher-level program logic over structure-preserving function-level rewrites.

\subsubsection{Data-Limited Fine-Tuning}
\label{sec:finetuning}

We study the real-world scenario of adapting to new malware samples, which are scarce and costly to label, by finetuning a pretrained GIN model on increasingly small subsets of the labeled OOD training data. We run this experiment over 3 seeds, as fewer data and training epochs can lead to higher variance, and plotting them in \Cref{fig:ft-truncate}.

\paragraph{Pre-Training Robustness.}
\label{sec:ptr}
We pretrained the model with different augmentation strategies, then run standard finetuning on the resulting checkpoints: in this setting, augmentations help the model learn a more robust representation space that can better adapt to the downstream distribution shifts upon encountering just a few new samples. The results in \Cref{fig:aug-base} show that node dropping improves over the unaugmented baseline, confirming its capacity to generate diverse data for pretraining, though this advantage does not carry into finetuning, as the next section shows. Conversely, edge dropping instead falls below the baseline, and is useful only when a checkpoint is consumed immediately after augmented model fitting.
Notably, structure-only FCGA-S markedly outperforms all others especially in the low-data regime, underscoring the value of structural diversity for pretraining.

\paragraph{Post-Training Generalization.}
\label{sec:ptg}
We perform standard pre-training without augmentation, but apply different augmentation strategies during finetuning: in this setting, augmentations serve as a partial substitute for human annotation, enriching the limited new malware samples with diverse variants. The results in \Cref{fig:aug-ft} reveal the opposite ordering for context-agnostic approaches: node-drop-based methods now falls far below the baseline, by up to 12\%, and is clipped from the plot for legibility; edge dropping takes its place as the weakest strategy that still improves on the baseline. FCGA variants cluster much more tightly in this setting, but the model-aware variants remain consistently the strongest across every data fraction.

\subsubsection{Robustness Under LLM-Rewrite Attack}
\label{sec:llm-attack}
\begin{table}[t]
\small
\centering
\begin{tabularx}{\linewidth}{lYcc}
\toprule
Method & OOD & LLM (seen) & LLM (unseen) \\
\midrule
Baseline & 57.7\% & 48.0\% & 34.0\% \\
\midrule
NodeDrop & 58.5\% & 63.0\% & 39.0\% \\
EdgeDrop & 59.9\% & 53.0\% & 38.0\% \\
Node+Edge & 58.7\% & 58.0\% & 36.0\% \\
\midrule
Function & \underline{62.0\%} & 67.0\% & \underline{43.0\%} \\
... + Model & 61.3\% & 67.0\% & 42.0\% \\
\midrule
Perturb & 61.4\% & \underline{70.0\%} & 41.0\% \\
... + Model & 60.3\% & 63.0\% & 41.0\% \\
... + All & 61.9\% & \textbf{72.0\%} & \underline{43.0\%} \\
\midrule
Subgraph & 60.9\% & 66.0\% & 39.0\% \\
... + Model & \textbf{64.4\%} & \underline{70.0\%} & \textbf{44.0\%} \\
\bottomrule
\end{tabularx}
\caption{Accuracy under out-of-distribution malwares of pretrained GIN models across different augmentation methods. ``Seen''/``unseen'' denotes LLM-rewritten train/test samples, respectively. Higher is better, largest value in each column in bold, second largest underlined.}
\label{tab:llm_attack}
\vspace{-0.75ex}
\end{table}

For a better assessment of robustness against LLM-rewrite attacks, we additionally craft a direct attack using a different LLM to simulate a real-world scenario.
Reproducing LLMalMorph~\cite{akil2025llmalmorphfeasibilitygeneratingvariant}, we rewrote the training and test sets of the MNTF dataset with Phi-4-Mini-3.8B~\cite{microsoft2025phi4minitechnicalreportcompact},
augmenting the first 100 samples from the training set and 100 from the test set across \textit{all} nodes; where the augmented training set represents the attack conducted on known malicious code, and the augmented test set represents which that has not been seen in the wild.
The results, reported in \Cref{tab:llm_attack}, show high correlation between OOD evaluation and the LLM attack, supporting \textit{Common} as a useful proxy for this threat.
Across all settings, FCGA-S + Model consistently yields the best or near-best performance, with all other FCGA variants outperforming the blind-drop baselines. One final interesting data point: NodeDrop performs notably well under the LLM attack on seen data, as the model learning to be robust to sudden node changes can be helpful in a node-only attack.
Note that this experiment only serves as a held out cross-model LLM rewrite stress test motivated by a realistic attack setting; and we do not claim overall generalization to an arbitrary LLM-rewrite attack.
\section{Discussion: Fidelity vs. Diversity}
\label{sec:tradeoffs}
Our experiment results illustrate a very important (and perhaps obvious) tradeoff in designing a suitable augmentation method for different training modes; that is, fidelity is crucial for any last-mile finetuning, while diversity is important for generalizable pre-training. This is elaborated from the blind node/edge-drop results, where node-drop improves pre-training robustness (\textit{Pre}, \Cref{sec:ptr}a) but not post-training generalizability (\textit{Post}, \Cref{sec:ptg}b), and edge-drop vice versa. Node drop is a more aggressive way to achieve structural diversity, but this destructiveness hurts the low-data finetuning setting. Edge drop is much gentler as information redundancy allows the model to learn from the augmented data without being misled by the noisy structure, since source codes still include dropped function calls. However, this results in less diverse augmentations, limiting edge drop's overall effectiveness.
We refer readers to Appendix~\ref{app:discussions} for further analysis and accompanying experiments.

\paragraph{Selection: reliability vs. coverage.}
The node selection strategies in \Cref{sec:method} span a spectrum from static topological measures to model-dependent explainability methods.
In practice, static criteria are preferable when the model is early in training or computational budget is limited: during pretraining, the model is only learning basic representations, and thus diversity of augmentation can be more beneficial. In contrast, model-dependent criteria become more valuable once the model has learned meaningful representations and targeted augmentation of its decision boundary is most consequential.

This observation explains how the top-3 results for \textit{Post} are all model-aware, while the top performer for \textit{Pre} is FCGA-S \textit{without} model-awareness. The latter phenomenon is caused by model-aware selection only affecting the model decision and not necessarily its propagation, in turn not impacting model's computation enough for robustness.
FCGA-P is interesting as it takes the trading off reliability for coverage to the extreme, blind-augmenting everything, and thus is effective in the situation where both \textit{Pre} and \textit{Post} apply (i.e. direct evaluation of pre-training, \Cref{sec:pretraining}), but neither individually. This is further exacerbated by its semantic approximation during perturbation.

\paragraph{Augmentation: semanticity vs. approximation.}
FCGA-P eliminates the LLM inference cost of FCGA-F through a closed-form approach, but at the expense of semantic grounding. As a result, FCGA-P variants underperform other approaches in both \textit{Post} (except the Model variant, whose selection reliability offsets semantic approximation) and \textit{Pre} settings. Decision boundary theory explains the latter: the optimal separator hyperplane learned by diverse-but-blind data-only augmentation will not hold in the presence of true distribution-shifted samples 
(see \Cref{fig:boundary} in Appendix~\ref{app:boundary}). 
Therefore, similar to the selection strategies, FCGA-P should only be used for cheap pre-training coverage with direct evaluation, while FCGA-F is reserved for fine-tuning stages where the model is sensitive to semantically meaningful variation.
This hybrid amortizes LLM inference cost while preserving the benefits of semantic diversity where they matter most.
Meanwhile, subgraph-based augmentation is effective across all settings, benefiting from both the augmentation diversity induced by structural edits and semantic refactoring.
The top-2 results for \textit{Pre} both belong to FCGA-S; but as powerful as it is, nothing good comes for free (Appendix~\ref{app:fcga-s}).
\section{Conclusion}
In this work, we studied the brittleness of FCG-based malware classifiers against LLM-rewritten variants.
We proposed FCGA, an LLM-driven augmentation framework that rewrites individual functions and refactors function clusters into realistic, semantically-diverse training variants, concentrating its effort on the graph components most important to the target model.
Across multiple GNN architectures and distribution-shift settings, including a direct LLM-rewrite attack, FCGA improves robustness by up to 10\% over context-agnostic baselines.
Beyond these gains, our analysis surfaces a fidelity-diversity tradeoff that dictates when semantic grounding and model-aware selection matter most across the training pipeline.
We hope this work encourages a more proactive defense against generative threats, leveraging the attacker's own strategy to strengthen the classifiers they target.

\section*{Limitations}

\paragraph{LLM inference overhead.}
FCGA-F and FCGA-S require LLM inference over the entire training set.
Although this cost incurs only once and is amortized across all subsequent training runs, it grows linearly with dataset size and number of augmented functions, making it non-trivial for large-scale malware corpora.
FCGA-P eliminates this overhead at the cost of semantic grounding, and can be very effective for pretraining, but severely underperforms when adapting to distributionally-shifted data. Additionally, ablation over the number of augmented nodes (c.f. Appendix \ref{app:ablations}) shows diminishing returns beyond a small augmentation budget, suggesting that even a limited number of LLM rewrites can provide significant robustness benefits.

\paragraph{Small language models.}
We employed Qwen2.5-Coder-7B~\cite{qwen2025qwen25technicalreport} as the code reasoner as it is the most accessible code model amongst the available open-source ones. Our preliminary test with Llama3.1-8B~\cite{grattafiori2024llama3herdmodels}
and Phi3-3.8B~\cite{abdin2024phi3technicalreporthighly}
yields inconsistent outputs in both reasoning and formatting, resulting in low-quality refactor outputs; and for more recent modern reasoning models, additional ablations can be found in \Cref{app:llm-ablations}. We expect our framework to perform even better with larger models and/or state-of-the-art proprietary models such as Claude and GPT-5; however, guardrailed models rejecting certain prompts deemed to be harmful hinders our ability to fully leverage their capabilities, and jailbreaking them falls outside the scope of this work. We defer this extensive benchmarking to a future work, particularly as more capable open-weight alternatives become available.

\paragraph{Android and Java specificity.}
All experiments are conducted on Android malware FCGs derived from decompiled Java bytecode.
While it's straightforward to apply FCGA to other platforms, the lack of benchmarks with FCG representations and available source code for other domains prevents direct evaluation of our framework's effectiveness beyond this setting.
Extending FCGA to other executable formats, e.g. Windows PE, Linux ELF, or native ARM binaries, is thus left for a future work.

\paragraph{Iterative augmentation.}
We currently augment the training data in a single pass, using a snapshot of the model state as the basis for selection criteria.
Iteratively augmenting the training set in multiple rounds, where the selection criteria are updated after every epoch similar to adversarial training~\cite{madry2019deeplearningmodelsresistant}, could potentially further enhance robustness by continuously adapting to the model's evolving weaknesses.
However, this would significantly increase the computational cost and complexity of the training process; meanwhile, even with an one-time augmentation, we reach near optimal accuracy with just 25\% of the training data, suggesting that the marginal benefit of iterative augmentation may not justify its cost.

\section*{Ethical Considerations}

\paragraph{Dual-use considerations.}
FCGA repurposes a capability that is already established and publicly documented: prior work has shown that LLMs can rewrite malware into evasive variants at scale~\cite{Madani_2023, akil2025llmalmorphfeasibilitygeneratingvariant, hortea2026infinitemutationenginemeasuring}.
Our framework introduces no new offensive technique and does not lower the barrier to producing malware; it instead operates entirely on the defensive side, consuming such variants as training signal to harden classifiers rather than synthesizing deployable malicious artifacts.
We contend that this defensive framing is necessary rather than optional given that adversaries already wield LLMs, and thus withholding the same tools from defenders only widens an existing asymmetry.
To this end, we release our code and prompts to support reproducibility and defensive adoption; the prompts specify generic, semantics-preserving refactoring rather than any malware-specific weaponization, and we believe their value to defenders outweighs the marginal informational risk of describing a technique already available to attackers.

\paragraph{Responsible data handling.}
All experiments use MalNet~\cite{freitas2021malnet} and its feature-augmented variant~\cite{mntf}, all under CC-BY-4.0 license, established public benchmarks of FCGs derived from decompiled Android applications and not live executables.
At no point do we execute, compile, or repackage malware into runnable form: FCGA operates over graph and embedding representations, and its augmentations are abstract training samples that need not be executable by design.
We likewise do not jailbreak or adversarially circumvent model safety mechanisms: the task we prompt for is semantics-preserving refactoring of existing functions, and we rely on open-weight code models run locally whose lighter alignment permits such code-transformation requests.

\section*{Acknowledgments}
This material is supported by the National Science Foundation under Award Numbers 2325416, 2325417, 2524380, 2540765, and 2622415.

\bibliography{ref}


\appendix
\section*{\Large Appendix}

\begin{table}[b]
\small
\centering
\begin{tabularx}{\linewidth}{Xcccc}
\toprule
Model & GCN & GIN & GAT & GraphSAGE \\
\midrule
\#Params. & 82,949 & 95,429 & 83,333 & 95,237 \\
\bottomrule
\end{tabularx}
\caption{Parameter counts of model architectures used in the main experiments.}
\label{tab:model_params}
\end{table}

\section{Implementation Details}
\label{app:implementation}

\subsection{Hardware \& Frameworks}
All experiments are conducted on a single NVIDIA RTX 6000 Ada Generation GPU.
GNN models are implemented and evaluated in PyTorch~2.7 and PyTorch Geometric~(PyG)~2.4.
LLM inference for both function-level rewriting and subgraph-level refactoring variants of FCGA uses Qwen2.5-Coder-7B-Instruct~\cite{qwen2025qwen25technicalreport} served locally via Ollama for results in the main paper;
additional LLM ablations are specified in their respective descriptions.
Node features are 1024-dimensional embeddings of their corresponding source codes produced by
Qwen3-Embedding-0.6B~\cite{qwen3technicalreport}, consistent with the setup of the MalNet-Tiny-Feature benchmark~\cite{mntf}.

\subsection{Training Hyperparameters}
Each architecture has 3 layers of message-passing, and a 2-layer MLP classification head, with model parameter counts listed in \Cref{tab:model_params}. MalNet-Tiny~\cite{freitas2021malnet} and MalNet-Tiny-Features~\cite{mntf} use identical splits with 3,500 training samples, 500 validation, and 1,000 test samples. All models are pretrained for 150 epochs with early stopping based on validation accuracy, using the Adam optimizer with a learning rate of 0.001 and weight decay of 5e-4. Batch size is set to 128 for all experiments. Finetuning is run for 10 epochs.
All experiments are seeded with 0 for reproducibility.
For data-limited fine-tuning, subsets of the training data are identically truncated by the specified ratio, evenly distributed across each class.

\subsection{Cosine Perturbation}
\label{app:cosine}

FCGA-P augments node embeddings by adding a scaled random noise vector whose direction is chosen uniformly at random and whose magnitude is calibrated so that the resulting embedding lands at a prescribed cosine similarity $c$ from the original.
$c$ is calibrated to match the empirical distributional shift of FCGA-F, ensuring that the two methods induce a comparable deviation in feature space.
We derive the closed-form scale below.

Let $\bm{a} \in \mathbb{R}^d$ be the original embedding and $\hat{\bm{a}} = \bm{a}/\|\bm{a}\|$ its unit normalization.
Sample $\bm{u} \sim \mathcal{N}(\bm{0}, \bm{I}_d)$, set $\hat{\bm{u}} = \bm{u}/\|\bm{u}\|$, and let us start with the noise vector $\bm{u}'=\Vert\bm{a}\Vert\hat{\bm{u}}$ for cleanliness of proof.
We seek a scalar $k$ such that
\[
  \cos\bigl(\bm{a},\; \bm{a} + k{\bm{u}'}\bigr)
  = \frac{\bm{a} \cdot (\bm{a} + k\Vert\bm{a}\Vert\hat{\bm{u}})}
  {\|\bm{a}\| \left\|\bm{a} + k\Vert\bm{a}\Vert\hat{\bm{u}}\right\|}
  = c.
\]
Define $b = \hat{\bm{a}} \cdot \hat{\bm{u}}$, the projection of the unit embedding onto the noise direction.
Dividing numerator and denominator by $\|\bm{a}\|$ and substituting $\|\hat{\bm{a}} + k\hat{\bm{u}}\|^2 = 1 + 2kb + k^2$ (since both vectors are unit-length) gives
\[
  \frac{1 + kb}{\sqrt{1 + 2kb + k^2}} = c.
\]
Squaring both sides (valid when $1 + kb > 0$, which holds for small $|b|$ and $c > 0$) and rearranging yields the quadratic
\begin{equation}
  (c^2 - b^2)\,k^2 + 2b(c^2 - 1)\,k + (c^2 - 1) = 0.
  \label{eq:cosine-quad}
\end{equation}
Its discriminant simplifies to
\begin{align*}
  \Delta
  & = 4b^2(c^2-1)^2 - 4(c^2 - b^2)(c^2 - 1) \\
  & = 4c^2\,(1-c^2)\,(1-b^2),
\end{align*}
which is strictly positive for $c \in (0,1)$ and $|b| < 1$, guaranteeing two distinct real roots:
\begin{equation}
  k_{\pm}
  = \frac{b(1-c^2) \pm c\sqrt{(1-c^2)(1-b^2)}}{c^2 - b^2}.
  \label{eq:cosine-roots}
\end{equation}
In $d$-dimensional space the projection $b$ concentrates near zero as $d$ grows large, so $b^2 \ll c^2$ with overwhelming probability.
When $c^2 > b^2$ the denominator is positive and $k_+ > 0 > k_-$; we therefore take $k^* = k_+$.
The perturbed embedding is
\begin{equation}
\label{eq:calc_cosine}
  \tilde{\bm{a}} = \bm{a} + k^*{\bm{u}'},
\end{equation}
which satisfies $\cos(\bm{a}, \tilde{\bm{a}}) = c$ by construction.
In the rare edge case $b^2 \geq c^2$, we re-sample $\bm{u}$.

The monotonicity of the cosine in $k$ can be verified by differentiating $c^2(k) = (1+kb)^2/(1+2kb+k^2)$ with respect to $k$:
\[
  \frac{\mathrm{d}c^2}{\mathrm{d}k}
  \;\propto\; k(b^2-1)(1+kb),
\]
which is negative for $k > 0$ and $|b| < 1$, confirming that the perturbed embedding moves strictly further from the original as $k$ increases.

\subsection{Node Selection}
\label{app:selection-criteria}

\subsubsection{Static Structural Importance}
Different centrality measures capture unique structural properties of a network, defining ``importance'' differently based on distinct types of network positioning or flow.
Degree centrality ranks nodes by the number of incident edges; high-degree functions are called frequently and act as hubs in the program's call structure.
Betweenness measures the fraction of shortest paths passing through a node, identifying bottleneck functions whose removal would disconnect large parts of the graph.
Closeness captures global centrality as the inverse average shortest-path distance to all other nodes.
PageRank recursively propagates importance through the graph, assigning higher scores to nodes reachable from many important callers.

Improving upon PageRank, we utilized Personalized PageRank (PPR) to the Android setting. Our PPR implementation seeds the random walk at Android lifecycle entry points such as \texttt{onCreate}, \texttt{onReceive}, and \texttt{onStartCommand}, the primary interfaces through which the Android runtime interacts with an application. Random walks have previously been widely used for graph learning~\cite{lao2011random,nikolentzos2020random}; and in the context of FCGs, they have a natural interpretation as simulating the flow of execution through the program, and thus was utilized as a surrogate for dynamic execution traces~\cite{onwuzurike2019mamadroid}.
By biasing selection toward functions reachable from these entry points, PPR surfaces nodes that lie on the execution paths most relevant to the application's observable behavior.

\subsubsection{Model-Aware Importance}
The simplest approach for model-aware node importance calculation is to measure the output performance drop when a graph component is removed. However, prominent methods in explainable AI more sophisticatedly analyze the statistical correlation between graph components and the model's prediction. GNNExplainer~\cite{ying2019gnnexplainergeneratingexplanationsgraph} determines importance by maximizing mutual information between graph nodes/edges and the model's prediction.
PGMExplainer~\cite{vu2020pgmexplainerprobabilisticgraphicalmodel} perturbs node features in batches and uses $\chi^2$ independence tests to measure which nodes' perturbations most shift the prediction.

On subgraph selection, SubgraphX~\cite{yuan2021explainability} finds important subgraphs utilizing MCTS with Shapley values~\cite{lundberg2017unified} as the reward signal, which attribute the model's output to each subgraph by computing its weighted average marginal contribution across all possible subsets.
While their choice of Shapley values is motivated by its theoretical properties, it is computationally expensive to evaluate even with stochastic approximation, requiring a nested Monte Carlo sampling routine for each MCTS state evaluation.
We further simplify this approach and experiment with replacing the expensive Shapley value computation with output drop in \Cref{app:model-criteria}, a simpler metric that is core to all explainability metrics, and can be evaluated efficiently in batches. The result of this process is a set of important subgraphs that are not only structurally cohesive, but also empirically influential to the model's predictions.

\subsection{PGMExplainer Implementation}
We reimplemented PGMExplainer with the following modifications to the current version on PyG Contrib to improve efficiency and correctness:
\begin{enumerate}
  \item Batched perturbation inference:
  The reference implementation runs \texttt{num\_samples} sequential forward passes per graph, with one model call per perturbed copy. We generate all \texttt{num\_samples} perturbed copies upfront, stack them into a single PyG Batch, and execute one batched forward pass per round. This reduces model calls from $O(\texttt{num\_samples})$ to $O(1)$ per round per graph.

  \item Round 1 chi-square target:
  In the reference, the chi-square target for Round 1 is \texttt{int(target)}, where target is the predicted class index passed by the Explainer wrapper. This integer is used as a column index into the sample matrix, whose columns record per-node perturbation flags (independent Bernoulli draws). Consequently, Round 1 tests whether node $k$'s correlates with node $\hat{y}$'s perturbation (i.e. two independent random variables), yielding p-values that are statistically uniform and carry no information about prediction sensitivity. We use \texttt{num\_nodes} (the binarized prediction-change column) as the chi-square target for both rounds, so Round 1 performs a genuine coarse filter over all nodes before Round 2 refines the candidate set.

  \item Candidate-set size for small graphs:
  The reference sets $\texttt{max\_subgraph\_size} = int(\texttt{num\_nodes} / 20)$ when no value is provided, which evaluates to zero for graphs with fewer than 20 nodes, producing an empty candidate set and degenerate Round 2 behavior. We compute $\texttt{max\_subgraph\_size} = max(2k,\lfloor \texttt{num\_nodes} / 20 \rfloor)$, where $k$ is the node budget, clamped to $[1,\texttt{num\_nodes} - 1]$, ensuring a non-empty candidate set.

  \item Gradient suppression:
  The reference runs all perturbation forward passes with autograd enabled. We wrap all model calls in \texttt{torch.no\_grad()}, which is functionally equivalent since PGMExplainer does not use gradients, to reduce memory overhead.
\end{enumerate}

\section{Ablation Studies}
\label{app:ablations}
In this section, we present detailed ablation studies over the effect of different hyperparameters for each augmentation method, and detailed the final configurations which are used in the experiments reported in the main text in \Cref{tab:hyperparameters}.
\begin{table}[t]
\small
\centering
\begin{tabularx}{\linewidth}{lYc}
\toprule
Method & Selection & Aug. Rate \\
\midrule
NodeDrop & Uniform & 0.25 \\
EdgeDrop & Uniform & 0.1 \\
Node+Edge & Uniform & (0.1, 0.5) \\
\midrule
Function & Betweenness & 0.8 \\
... + Model & PGMExplainer & 0.2 \\
\midrule
Perturb & PPR & 0.5 \\
... + Model & PGMExplainer & 0.5 \\
... + All & All & 0.2 \\
\midrule
Subgraph & Degree & 0.2 \\
... + Model & SubgraphX & 0.2 \\
\bottomrule
\end{tabularx}
\caption{Final hyperparameter selections for each augmentation methods used in the main experiments.}
\label{tab:hyperparameters}
\end{table}

\begin{table*}[t]
\centering
\small
\begin{tabularx}{\textwidth}{cYYYYYYYY}
\toprule
Edge Rate $\rightarrow$ & \multicolumn{2}{c}{0} & \multicolumn{2}{c}{0.1} & \multicolumn{2}{c}{0.3} & \multicolumn{2}{c}{0.5} \\
\cmidrule(lr){2-3}\cmidrule(lr){4-5}\cmidrule(lr){6-7}\cmidrule(lr){8-9}
$\downarrow$ Node Rate &
Test & OOD & Test & OOD & Test & OOD & Test & OOD  \\
\midrule
0    & 94.7\% & 57.7\% & 95.0\% & \underline{59.9\%} & \underline{95.4\%} & 59.0\% & 94.8\% & 59.3\% \\
\midrule
0.05 & 94.4\% & 56.5\% & 95.4\% & 56.7\% & 95.2\% & 57.7\% & 95.2\% & 58.1\% \\
0.10 & 94.9\% & 56.4\% & \textbf{95.6\%} & 57.9\% & 93.9\% & 55.4\% & 94.9\% & \textbf{58.7\%} \\
0.25 & \underline{95.0\%} & \underline{58.5\%} & 94.8\% & 57.6\% & 95.4\% & 57.7\% & 95.3\% & 58.6\% \\
\bottomrule
\end{tabularx}
\caption{Prediction accuracy on the original and covariance-shifted MalNet-Tiny datasets across node and edge drop rates. Trained on the GIN architecture. Higher is better. Highest value in Edge/Node-only ablations underlined, highest value across Node+Edge in bold.}
\label{tab:drop_ablation}
\end{table*}

\subsection{Node Drop \& Edge Drop}
\label{app:drop-ablation}

We sweep drop rates $r_n \in \{0.05, 0.10, 0.25\}$ for NodeDrop and $r_e \in \{0.1, 0.3, 0.5\}$ for EdgeDrop jointly on GIN, measuring both \textit{Test} and \textit{OOD} accuracy.
\Cref{tab:drop_ablation} reports all configurations, with the first row ($r_n = 0$) isolates EdgeDrop, the first column ($r_e = 0$) isolates NodeDrop, and $r_n = r_e = 0$ represents the baseline.

Test accuracy is stable across configurations, staying within roughly 2\% of the baseline.
OOD accuracy is more sensitive: EdgeDrop alone with $r_e = 0.1$ achieves 59.9\%, while NodeDrop alone with $r_n = 0.25$ yields 58.5\% with consistent gains on both splits.
The combined Node+Edge setting peaks at $r_n = 0.10, r_e = 0.5$ for OOD with 58.7\%.

\subsection{Graph Component Selection Criteria}

\begin{table}[!b]
\small
\renewcommand{\arraystretch}{0.9}
\centering
\begin{tabularx}{\linewidth}{p{1.3cm}cYYY}
\toprule
{Criteria} & {$\mathbb{E}[\#n]$} & {FCGA-F} & {FCGA-P} & {FCGA-S} \\
\midrule
Baseline & - & 58.0\% & 58.0\% & 58.0\% \\
\midrule
\multirow{4}{*}{Degree}      & 1 & \underline{57.3\%} & \underline{60.6\%} & 58.2\% \\
& 2.5 & 53.8\% & 57.0\% & 60.8\% \\
& 4 & 56.8\% & 57.6\% & 58.9\% \\
& 5 & 56.9\% & 60.2\% & \underline{60.9\%} \\
\midrule
\multirow{4}{*}{Betweenness} & 1 & 60.2\% & \underline{60.4\%} & \underline{59.0\%} \\
    & 2.5 & 58.6\% & 56.8\% & 58.0\% \\
    & 4 & \textbf{{62.0\%}} & 59.7\% & 58.6\% \\
    & 5 & 58.7\% & 59.1\% & 56.7\%\\
\midrule
\multirow{4}{*}{Closeness}   & 1 & \underline{61.4\%} & \underline{{57.3\%}} & 58.6\% \\
    & 2.5 & 57.7\% & 57.0\% & 58.3\% \\
    & 4 & 59.2\% & 56.5\% & 58.5\% \\
    & 5 & 58.3\% & 57.1\% & \underbar{59.3\%} \\
\midrule
\multirow{4}{*}{PPR}         & 1 & 60.0\% & 59.9\% & 58.1\% \\
    & 2.5 & \underline{60.3\%} & \underline{61.4\%} & 58.8\% \\
    & 4 & 55.7\% & 59.7\% & \underline{59.0\%} \\
    & 5 & 59.8\% & 56.9\% & 56.7\% \\
\midrule
\multirowcell{4}[0pt][l]{GNN\\Explainer}         & 1 & 59.3\% & 57.9\% & - \\
    & 2.5 & 59.7\% & 57.9\% & - \\
    & 4 & \underline{59.8\%} & \underline{59.0\%} & - \\
    & 5 & 58.2\% & 58.7\% & - \\
\midrule
\multirowcell{4}[0pt][l]{PGM\\Explainer}         & 1 &  \underline{61.3\%} & 58.8\% & - \\
    & 2.5 & 58.4\% & \underline{60.3\%} & - \\
    & 4 & 58.3\% & 58.2\% & - \\
    & 5 & 58.4\% & 58.0\% & - \\
\midrule
\multirowcell{4}[0pt][l]{MCTS\\+ Output}         & 1 & - & - & 59.1\% \\
    & 2.5 & - & - & 56.4\% \\
    & 4 & - & - & \underline{59.2\%} \\
    & 5 & - & - & 58.9\% \\
\midrule
\multirow{4}{*}{SubgraphX}         & 1 & - & - & \textbf{64.4\%} \\
    & 2.5 & - & - & 60.5\% \\
    & 4 & - & - & 59.3\% \\
    & 5 & - & - & 58.9\% \\
\midrule
\multirow{4}{*}{All}         & 1 & - & \textbf{61.9\%} & - \\
    & 2.5 & - & 59.2\% & - \\
    & 4 & - & 57.7\% & - \\
    & 5 & - & 56.9\% & - \\
\bottomrule
\end{tabularx}
\caption{Prediction accuracy on the covariance-shifted MalNet-Tiny datasets across different node sampling criteria and augmentation rates. Trained on the GIN architecture. Higher is better. Largest value in each column in bold, largest value for each node selection criteria underlined.}
\label{tab:node_criteria}
\end{table}
We jointly ablate the node selection criterion $\in \{\text{Degree, Betweenness, Closeness, PPR}\}$ and the expected number of augmented nodes per graph $\mathbb{E}[\#n] \in \{1, 2.5, 4, 5\}$ for both FCGA-F and FCGA-P on GIN. This expectation is implemented by augmenting the most important 5 nodes and individually sampling a random Bernoulli variable for each of them with probability $p\in\{0,0.2,0.5,0.8,1.0\}$ across the training set.
We report the results in \Cref{tab:node_criteria}.

\subsubsection{Model Performance}
\paragraph{Static topology-based criteria.}\label{app:static-criteria}
With function-level augmentation, similarly to the drop ablation, Test accuracy is relatively stable across configurations, with a maximum spread of around 1\%, while OOD accuracy varies more widely with hyperparameters.
Specifically, FCGA-F consistently achieves above 60\% OOD accuracy across all criteria except Degree, and FCGA-P achieves similar performance but except for closeness. Interestingly, the trend for FCGA-S is the complete opposite of FCGA-F, where only Degree-based selection achieves above 60\% OOD accuracy. This suggests that fancier centralities criteria may not be well-aligned with MCTS in the malware FCG domain. Another interesting note is that FCGA-F with only $\mathbb{E}[\#n]=4$ edges over all-node FCGA-P, showing that while mass perturbation is very effective, concentrating high-quality augmentation on a limited set of high-impact nodes is still more beneficial in improving model robustness.

\paragraph{Model-dependent criteria.} \label{app:model-criteria}
For FCGA-F, PGMExplainer yields the best result at 61.3\%, however still trailing behind the best static selection method by 0.4\%, indicating that augmenting high-throughput nodes matters more than high model-attention. As they affected propagation the most, these nodes when perturbed should introduce more helpful variations in the representations.
For FCGA-P, a similar trend appears, where PGMExplainer outperforms GNNExplainer in improving OOD accuracy, but still cannot beat the performance of its own static selection counterparts.
Regarding FCGA-S, output-drop-guided MCTS disappointingly achieves sub-60\% OOD accuracy, while SubgraphX performs vastly better than any other configuration in the table. This implies output-drop as a reward proxy for the more computationally-expensive Shapley values may not be effective in this domain, which is consistent with how it's not commonly used in the explainer literature. Moreover, contrasting the effectiveness of static node selectors for function-level rewriting, this observation also reinforces the unsuitability of topology-based subgraph selection for augmentation.

\begin{table}[t]
\small
\centering
\begin{tabularx}{\linewidth}{lYcc}
\toprule
Config & Qwen2.5-Coder & Phi4-Mini & Gemma 4 \\
\midrule
FCGA-F & 60.3\% & \textbf{63.7\%} & 61.4\% \\
FCGA-S & 64.4\% & \textbf{64.6\%} & 62.0\% \\
\bottomrule
\end{tabularx}
\caption{OOD test set accuracy when using different augmenting LLMs. Higher is better, highest in bold.}
\label{tab:llm_ablation}
\end{table}

\subsection{Augmentation Count}
Two observations guide our hyperparameter selection in \Cref{tab:node_criteria}:
first, the selection criterion matters more than the augmentation count: varying $\mathbb{E}[\#n]$ at a fixed criterion produces modest OOD swings of 3-5\%, while the criterion decides the distribution of those gains.
Betweenness centrality with $\mathbb{E}[\#n] = 4$ gives the best FCGA-F configuration with OOD accuracy of 62.0\%, and Closeness with $\mathbb{E}[\#n] = 1$ gives the best FCGA-P OOD at 61.4\%.
Second, small augmentation counts suffice: configurations with $\mathbb{E}[\#n] = 1$ are largely competitive with larger counts across all criteria — for instance, Betweenness at $\mathbb{E}[\#n] = 1$ achieves 95.3\% Test and 60.2\% OOD, only 1.8\% below the best OOD at $\mathbb{E}[\#n] = 4$. As analyzed in \Cref{app:static-criteria}, this is thanks to centralizing augmentation around nodes that spread information throughout the graph, maximally impacting the underlying representation across the whole graph after model propagation.
This justifies using a small count in practice, which significantly reduces LLM inference time and cost.

\vspace{-0.25ex}
\section{In-Depth Analysis and Discussions}
\label{app:discussions}
\vspace{-0.25ex}

\subsection{LLM Ablations}
\label{app:llm-ablations}
\vspace{-0.25ex}

To assess the sensitivity of FCGA-F to the choice of LLM, we experiment with using Qwen2.5-Coder-7B~\cite{qwen2025qwen25technicalreport}, Phi4-Mini-3.8B~\cite{microsoft2025phi4minitechnicalreportcompact}, and Gemma 4-E4B~\cite{GoogleDeepMind2026Gemma4} as the augmenting LLM. For FCGA-F, we use PPR as it achieved consistently high result for both FCGA-F/P; and for FCGA-S, we select SubgraphX for its overall leading performance. Reported in \Cref{tab:llm_ablation}, results show that a stronger reasoning LLM can indeed yield better augmentation quality and thus better robustness, up to 3.4\% increase in accuracy.
Phi4-mini displays improved performance in both settings, although marginally for FCGA-S; while Gemma 4 was better than Qwen2.5-Coder for FCGA-F, but worse for subgraph-level refactoring.
Thus, we believe that our framework can further benefit from more powerful LLMs, and the results reported in the main text with Qwen2.5-Coder can be viewed as a lower bound for the potential of our framework.

\begin{table}[t]
\small
\centering
\renewcommand{\arraystretch}{0.9}
\begin{tabularx}{\linewidth}{lYYY}
\toprule
Method & Original & Gemma 4 & Phi-4-mini \\
\midrule
Baseline & 94.7\% & 87.0\% & 34.0\% \\
\midrule
NodeDrop & 95.0\% & 90.0\% & 39.0\% \\
EdgeDrop & 95.0\% & 93.0\% & 38.0\% \\
Node+Edge & 94.9\% & 93.0\% & 36.0\% \\
\midrule
Function & 95.0\% & 94.0\% & \underline{43.0\%} \\
... + Model & 95.0\% & 94.0\% & 42.0\% \\
\midrule
Perturb & 95.0\% & 94.0\% & 41.0\% \\
... + Model & 94.7\% & 93.0\% & 41.0\% \\
... + All & \textbf{95.1\%} & \textbf{96.0\%} & \underline{43.0\%} \\
\midrule
Subgraph & 94.4\% & 94.0\% & 39.0\% \\
... + Model & 94.9\% & \underline{95.0\%} & \textbf{44.0\%} \\
\bottomrule
\end{tabularx}
\vspace{-0.5ex}
\caption{Accuracy under LLM-rewritten test-set malwares of pretrained GIN models across different augmentation methods, using different attack LLMs. Higher is better, largest value in each column in bold, second largest using a different LLM underlined.}
\label{tab:llm_attack_gemma4}
\vspace{-1.5ex}
\end{table}

We additionally run our real-world attack in \Cref{sec:llm-attack} using Gemma 4 as the underlying LLM engine, and report the results in \Cref{tab:llm_attack_gemma4}. We found that Gemma-rewritten malwares did not drastically reduce the predictive accuracy of the models comparing to Phi-4-based ones, and stay much closer to the which on original test set. Comparing to the original test, FCGA variants achieve similar results on the Gemma-rewritten malwares, showing the ability to generalize to attacks using unseen LLMs.

\subsection{Selection Specificity Analysis}
\begin{figure*}[t]
    \centering
    \begin{subfigure}[t]{0.49\linewidth}
        \centering
        \includegraphics[width=\linewidth]{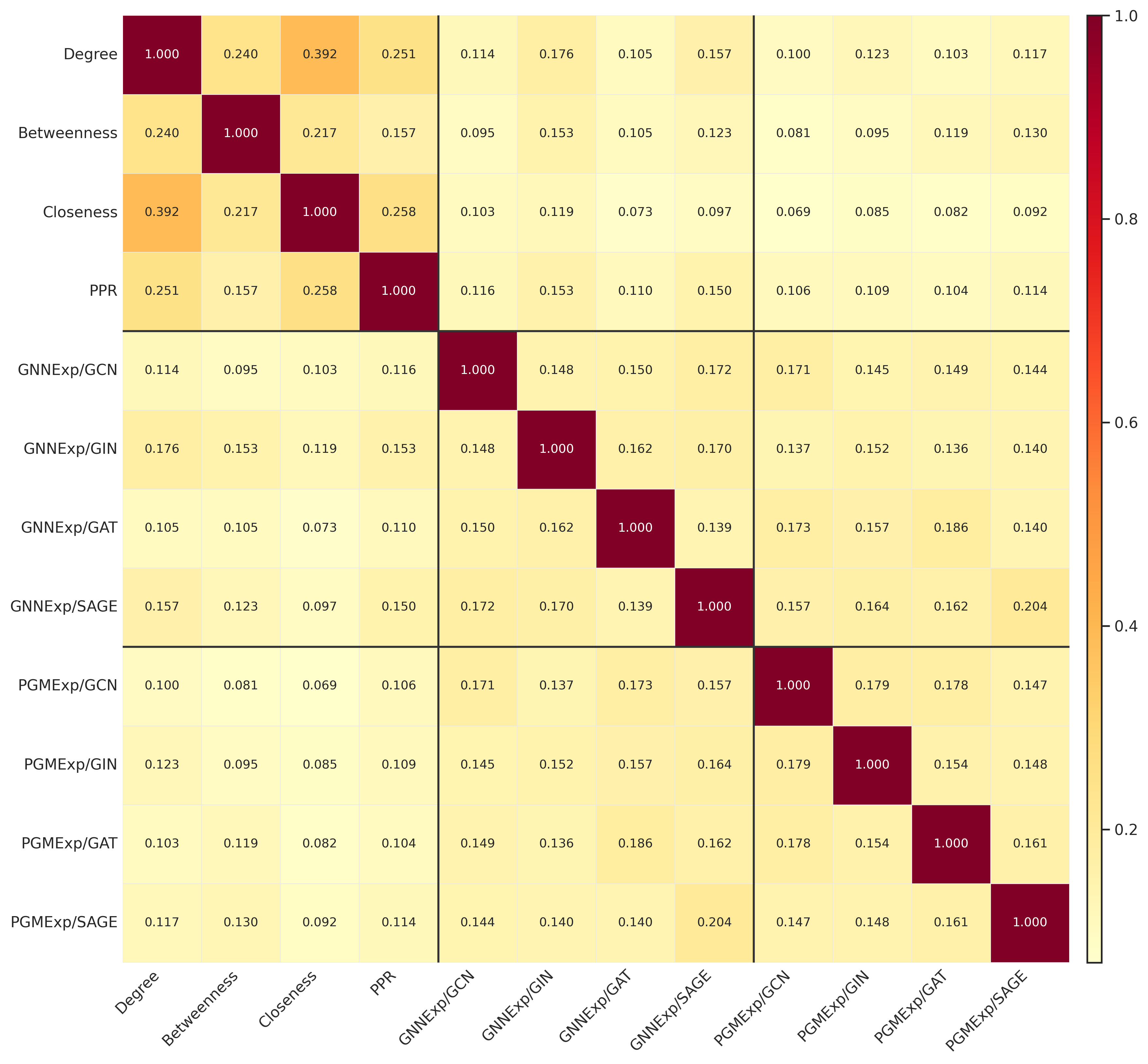}
        \caption{Function-Level}
    \end{subfigure}%
    ~ 
    \begin{subfigure}[t]{0.49\linewidth}
        \centering
        \includegraphics[width=\linewidth]{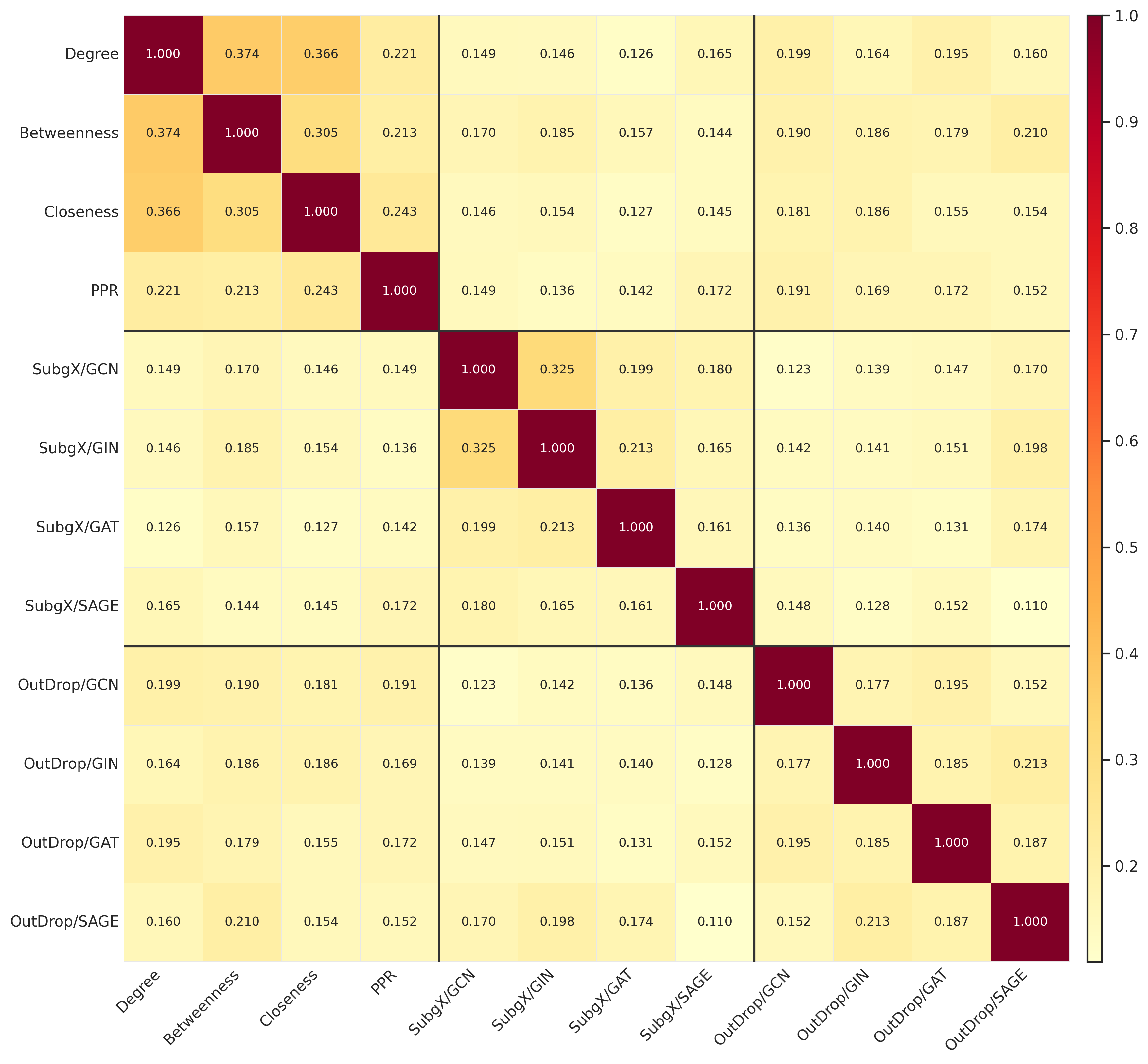}
        \caption{Subgraph-Level}
    \end{subfigure}
    \caption{Average Intersection-over-Union of node selections across criteria and checkpoints.}
    \label{fig:selection-iou}
\end{figure*}

\Cref{fig:selection-iou} visualizes the average overlap between selection criteria evaluated on the training set using intersection-over-union (IoU). Across the two FCGA granularities, static topology-based criteria would have more overlap comparing to vs. model-aware variants, but do not exceed 0.4 IoU. And vice versa, explainer-based selectors have more overlap with each other than with static ones, but much less than which across static criteria, evident by the general color intensities of the $4\times 4$ patches. The heatmap also shows that selection across architectures vastly differ from each other, denoting that each architecture learns a different set of classification rules, and thus model-dependent importance selection do not transfer well across setups.

\subsection{False Positive Rate on Benign Samples}
One possible concern about high-diversity augmentations is that their resulting aggressive expansion of the decision boundary to capture more malware variants (especially via FCGA-P) could lead to the over-rejection of benign code, which is critical for a real-world malware detector. However, this concern is alleviated by the fact that MNTF includes benign applications as a class in the training set, and thus these benign training samples constrain the decision boundary expansion during training. To inspect this, we evaluated FPR of the benign class (i.e. benign samples predicted as any malware family) of the experiments reported in Table~\ref{tab:pretrain} on the OOD split, and report the results in Table~\ref{tab:fpr}.

\begin{table}[t]
\small
\centering
\begin{tabularx}{\linewidth}{lYYYc}
\toprule
Method & GCN & GIN & GAT & GraphSAGE \\
\midrule
Baseline & 0.09 & 0.085 & 0.105 & 0.12 \\
\midrule
NodeDrop & 0.065 & 0.055 & 0.095 & 0.1 \\
EdgeDrop & 0.085 & \underline{0.095} & 0.095 & 0.08 \\
Node+Edge & 0.08 & 0.07 & \underline{0.22} & 0.11 \\
\midrule
FCGA-F & 0.065 & \textbf{0.045} & \underline{0.115} & \textbf{0.045} \\
... + Model & 0.06 & 0.065 & 0.05 & 0.065 \\
\midrule
FCGA-P & 0.085 & 0.08 & 0.05 & 0.09 \\
... + Model & 0.065 & 0.085 & \textbf{0.04} & 0.1 \\
... + All & 0.06 & 0.075 & 0.07 & 0.095 \\
\midrule
FCGA-S & 0.08 & 0.075 & \underline{0.11} & 0.09 \\
... + Model & \textbf{0.05} & 0.05 & \underline{0.11} & 0.075 \\
\bottomrule
\end{tabularx}
\caption{FPR on test set accuracy across different baselines and architectures. Lower is better, lowest in bold, values larger than baseline underlined.}
\label{tab:fpr}
\end{table}

FCGA-P in fact never increased FPR relative to the corresponding baseline, and reduced GAT FPR from 10.5\% to as low as 4.0\%. An FCGA configuration also attained the lowest FPR for every architecture. Nevertheless, the effect is architecture-dependent, especially with GAT, where augmentations may increase FPR, most notably Node+Edge at 22.0\%.
Determining the cause of this GAT-specific interaction would require a separate controlled study of architecture–augmentation interactions and is outside the scope of the current work.

\subsection{LLM Generation Quality}
To evaluate the quality of the rewrite output, we first ran \texttt{tree-sitter} on all Qwen-2.5 rewrites to check for syntactic validity. Out of a total of 372,576 rewrites, 163 (0.04\%) contain invalid syntax; and out of those, 82 are due to context limit cutoff. We then went through the other 81 invalid examples: most of them are actually functionally sound, but not necessarily correct, with the exception of 5 that have:

\vspace{-0.5ex}
\begin{itemize}[itemsep=1pt]
    \item Repeated blocks of code,
    \item Hallucinated variable type (e.g. calling \texttt{hasNext} on a \texttt{String}),
    \item Garbled text (e.g. \texttt{ifidwifi}),
    \item Early termination/truncation,
    \item A comment of \texttt{// (rest of the code remains unchanged)}.
\end{itemize}

\vspace{-0.5ex}
For the other 76, we provide a list of syntactic error categories:

\vspace{-0.5ex}
\begin{itemize}[itemsep=1pt]
    \item Generic invalid syntax (e.g. missing semi-colons, missing/misplaced closing braces),
    \item Reference before definition,
    \item Out-of-scope variable reference,
    \item Mistyped assignment/definition,
    \item Incorrect access modifier, or invalid access to non-public members,
    \item Mixed Kotlin code with Java,
    \item Deprecated syntax (\texttt{\textunderscore} for method identifier, which works in Java 8, but is reserved and illegal in Java 9+).
\end{itemize}

\vspace{-0.5ex}
And semantic errors:

\vspace{-0.5ex}
\begin{itemize}[itemsep=1pt]
    \item Loop counter never incremented,
    \item Function call before initializing parameters,
    \item Incorrect lifecycle (\texttt{setContentView} and \texttt{findViewById} called in the constructor instead of \texttt{onCreate} setup).
\end{itemize}

\vspace{-0.5ex}
We randomly (with seed 42) spot checked another 100 functions of the valid ones, and the results break down to:

\vspace{-0.5ex}
\begin{itemize}[itemsep=1pt]
    \item 61 equivalent or near-equivalent re-implementations of the original functionality,
    \item 25 functions with the right intention but bugs that prevent them from working correctly, with similar failure modes,
    \item 14 functions deviated from the original; most of them starts around the half-way mark.
\end{itemize}

\vspace{-0.5ex}
Similarly for subgraph refactoring, we randomly sample 100 of 31,399 refactored subgraphs:

\vspace{-0.5ex}
\begin{itemize}[itemsep=1pt]
    \item 59 were equivalent or near-equivalent,
    \item 25 had the right intention, with some incorrect or empty summaries,
    \item 10 functionally deviated,
    \item 6 output unparsable responses.
\end{itemize}

In light of these results, to strengthen the quality of the evaluation under LLM attack, we filter out functions that cannot be parsed by \texttt{tree-sitter} for the results in Table~\ref{tab:llm_attack} and Table~\ref{tab:llm_attack_gemma4}.

\begin{figure}[t]
    \centering
    \includegraphics[width=0.97\linewidth]{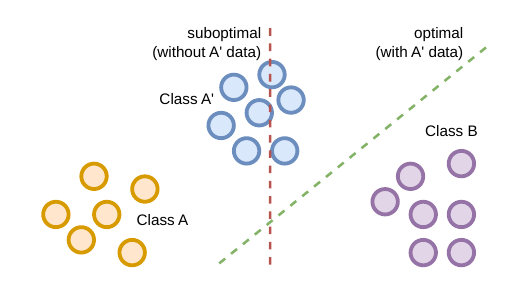}
    \caption{Illustration of the boundary induced by broad-but-blind FCGA-P, and the optimal boundary.}
    \label{fig:boundary}
    \vskip -1.5ex
\end{figure}

\subsection{Decision Boundary Intuition} \label{app:boundary} \Cref{fig:boundary} illustrates why broad but semantically ungrounded augmentation may yield a suboptimal decision boundary despite increasing training diversity, as it fails to cover the shifted distribution.

\subsection{The Cost of Thoroughness for FCGA-S}
\label{app:fcga-s}
Both components of FCGA-S require more computational overhead than their counterparts.
For selection, FCGA-S takes approximately 1 minute per graph on average, which is around the half the time needed for the LLM queries; while FCGA-F/P selection is much faster even with model-awareness. Given both MCTS and Shapley values are estimated via sampling, sufficient exploration is needed to avoid introducing variance that can undermine the reliability of selection.
We attempted to simplify this complexity by replacing Shapley values with output drop for a substantially reduction in runtime in~\Cref{app:model-criteria}; however the result is noticeably degraded, signaling that this computational overhead is highly justified. Regarding LLM inferences, while function-level rewriting only requires one prompt per augmented node and can be done in parallel, subgraph-level refactoring requires one prompt per node summarization node (which contains both augmenting nodes and context nodes), and another prompt for the refactoring itself that requires the completion of the summarization step. Moreover, while function rewrites need not be completely accurate, subgraph refactoring requires a valid JSON output for parsing and re-assembly. This requirement is more stringent and thus more prone to failure, especially with smaller models, which can further increase the overhead due to re-tries.
In our implementation, generated outputs are automatically schema-validated, and malformed outputs are retried under a fixed retry budget. Given that augmentation is performed offline, these retries affect one-time preprocessing cost rather than model training or inference cost.
Lastly, as mentioned in \Cref{sec:tradeoffs}, the complexity of subgraph refactoring demands a capable reasoning model with a sufficiently large context window (and consequently resistant to context rot). Thus, while FCGA-S yield visibly better robustness improvements, its computational cost and potential risks should be carefully considered in practice.

\begin{table}[t]
\small
\centering
\begin{tabularx}{\linewidth}{lYY}
\toprule
Variant & Test & OOD \\
\midrule
Baseline & 94.7\% & 57.7\% \\
\midrule
FCGA-F & \textbf{95.0\%} & 62.0\% \\
FCGA-P & \textbf{95.0\%} & 61.4\% \\
FCGA-Sum & 94.9\% & 58.4\% \\
FCGA-S & 94.4\% & 60.9\% \\
\midrule
FCGA-F + Model & \textbf{95.0\%} & 61.3\% \\
FCGA-P + Model & 94.7\% & 60.3\% \\
FCGA-Sum + Model & 94.8\% & 58.2\% \\
FCGA-S + Model & 94.9\% & \textbf{64.4\%} \\
\bottomrule
\end{tabularx}
\caption{Test set accuracy across different augmentation strategies on GIN. OOD refers to the \textit{Common} split. Higher is better, highest in bold.}
\label{tab:fcga-sum}
\vskip -2ex
\end{table}

Given the above limitations, FCGA-S currently combines its functional change from the LLM refactoring, and a representation change when embedding summaries instead of source code. To separate the impact of these two components to the effectiveness of FCGA-S, we run an addition ablation study where after subgraph selection and summarization, we do not run a refactor inference; and instead only embed those summarizations as features for the selected nodes. This results in an unchanged topology, but with the new summarization embeddings, eliminating one part of the confound. Denoting this new ablation as FCGA-Sum, we train a GIN model with this custom variant, and report the results on the OOD split in Table~\ref{tab:fcga-sum}. At 58.4\% (58.2\% with model-aware selection), FCGA-Sum barely improves OOD performance over baseline at 57.7\%, and worse than FCGA-F/P. This rules out that the effectiveness of FCGA-S stems mostly from augmenting source codes into summaries, and instead is due to the actual refactoring.
Unfortunately, the inverse ablation is practically infeasible: LLM's context limit prevents us from having both the refactor and new source codes. Thus, a complete separation of FCGA-S' structure-feature confound remains open for future research. However, our results with summary embeddings performing worse than which of actual function rewrites suggest that a complete refactor will perform even better than our current FCGA-S.

\section{Additional Related Work}
\subsection{Graph Representation for Code}

Representing programs as graphs is a fundamental design choice in learning-based code analysis. Multiple graph abstractions have been proposed, each offering a different tradeoff between expressiveness and graph size~\cite{allamanis2018survey}.

\textbf{Abstract Syntax Trees (ASTs)} capture the hierarchical syntactic structure of source code, with nodes corresponding to syntactic constructs and edges encoding parent-child relationships~\cite{alon2019code2vec}.
ASTs are expressive and widely used in code understanding tasks, but their size scales directly with code length, thus impractical for whole-program analysis of real-world applications.

\textbf{Control Flow Graphs (CFGs)} model the possible execution orders of basic blocks within a function, with edges representing transitions between them.
CFGs are well-suited for intra-procedural analysis such as vulnerability detection~\cite{zhou2019devign}, but remain scoped to individual functions and do not capture inter-procedural behavior.

\textbf{Data Flow Graphs (DFGs)} encode how values propagate through a program by connecting variable definitions to their uses~\cite{guo2021graphcodebert}.
Like CFGs, DFGs are typically constructed per-function and are often combined with CFGs for richer program representations.

\textbf{Code Property Graphs (CPGs)}~\cite{yamaguchi2014modeling} unify ASTs, CFGs, and DFGs into a single joint graph, simultaneously exposing syntactic structure, control flow, and data dependencies.
Despite their richness, CPGs are extremely large: a single real-world application can yield a graph with millions of nodes, making GNN-based inference computationally intractable at scale.

\textbf{Function Call Graphs (FCGs)} take a coarser, inter-procedural view: each node represents one function, and a directed edge from $v_i$ to $v_j$ indicates that function $v_i$ invokes $v_j$ during execution.
While FCGs do not expose the internal implementation of individual functions, they capture the high-level behavioral structure of a program, i.e. which components call which others, at a scale that remains tractable for GNN training.

\subsection{LLMs for Code Understanding}

Large language models pre-trained on large corpora of source code have demonstrated strong capabilities across a wide range of code-related tasks.
Early transformer-based models for code~\cite{feng2020codebert, wang2021codet5} established that jointly pre-training on natural language and programming languages enables effective code search, summarization, and understanding.
Subsequent generative models~\cite{chen2021codex, roziere2024codellama} further showed that LLMs can produce syntactically and semantically correct code from natural language descriptions, achieving strong performance on programming benchmarks.
These advances reflect a broader understanding that LLMs capture not only surface-level syntax but also the functional intent behind code.
Beyond generation, LLMs have proven effective at higher-level code reasoning tasks, including code summarization~\cite{koziolek2024llm}, refactoring assistance~\cite{nam2024usingllmhelpcode}, and bug localization~\cite{jimenez2024swebench}.


\begingroup
\makeatletter
\let\oldalgofinish\@algocf@finish
\renewcommand{\@algocf@finish}{%
    \oldalgofinish
    \par\vspace*{-2ex}%
}
\makeatother
\begin{algorithm}[t]
\small
\caption{FCGA-F: Function-Level Rewriting}
\label{alg:rewrite}
\KwIn{
    Graph $G=($ \\
    \Indp
    \Indp
    vertices $V=\{v_i\}$, \\
    edges $E$, \\
    node features $X=\{x_i\}, x_i\in\mathbb{R}^d$ \\
    \Indm
    $)$, \\
    source codes $S=\{s_i\}, s_i\in\mathcal{S}$, \\
    node selection $R=\{r_j\}, r_j\in 1..|V|$, \\
    rewriter LLM $L:\mathcal{S}\rightarrow\mathcal{S}$, \\
    embedding model $M:\mathcal{S}\rightarrow\mathbb{R}^d$
}
\KwOut{Augmented graph $G'$}
Initialize $X'\leftarrow X$\;
\ForEach{$r_j$ in $R$}{
    $X'[r_j]\leftarrow M(L(s_{r_j}))$\;
}
\Return{$G'=(V,E,X')$}\;
\end{algorithm}
\endgroup 

\subsection{LLMs for Graph Augmentation}

LLMs have more recently been brought into graph augmentation for text-attributed graphs (TAGs), where nodes carry natural-language descriptions.
TAPE~\cite{he2024harnessingexplanationsllmtolminterpreter} enriches node features with LLM-generated explanations, improving GNN performance on citation and product graphs.
\citet{sun2025large} prompt LLMs to assess pairwise semantic similarity between node texts and use the scores to add reliable edges and prune noisy ones, refining graph topology without changing node features.
\citet{yu2025leveraging} go further and generate entirely new labeled nodes via LLMs to address few-shot node classification on TAGs, integrating the synthetic nodes into the original graph through a learned edge predictor.
To the best of our knowledge, no prior work has applied LLM-driven augmentation to program graphs for malware detection.

\section{Algorithms and Prompts}

\subsection{Algorithms}

In this section, to accompany our code base and following prompts, we provide pseudo-code for our rewrite, perturbation, and refactoring algorithms at Algorithm~\ref{alg:rewrite}, ~\ref{alg:perturb}, and \ref{alg:refactor}, respectively.

\begingroup
\makeatletter
\let\oldalgofinish\@algocf@finish
\renewcommand{\@algocf@finish}{%
    \oldalgofinish
    \par\vspace*{-2ex}%
}
\makeatother
\begin{algorithm}[t]
\small
\caption{FCGA-P: Function-Level Perturbation}
\label{alg:perturb}
\KwIn{
    Graph $G=($ \\
    \Indp
    \Indp
    vertices $V=\{v_i\}$, \\
    edges $E$, \\
    node features $X=\{x_i\}, x_i\in\mathbb{R}^d$ \\
    \Indm
    $)$, \\
    cosine similarity $c\in\mathbb{R}$, \\
    node selection $R=\{r_j\}, r_j\in 1..|V|$
}
\KwOut{Augmented graph $G'$}
Initialize $\tilde{X}=X$\;
\ForEach{$r_j$ in $R$}{
    \tcp{Equation~\ref{eq:calc_cosine}}
    Generate $\tilde{x}_{r_j}$ s.t. $|\tilde{x}_{r_j}^\top{x}_{r_j}|/(\Vert\tilde{x}_{r_j}\Vert\Vert{x}_{r_j}\Vert) = c$\; 
    $\tilde{X}[r_j]\leftarrow \tilde{x}_{r_j}$\;
}
\Return{$G'=(V,E,\tilde{X})$}\;
\end{algorithm}
\endgroup

\subsection{Prompts}
\label{app:prompts}
\subsubsection{Function-Level Rewriting}
\begin{lstlisting}[style=promptstyle]
You are a Java code rewriter. Your task is to rewrite the following Java function to be as different as possible while STRICTLY preserving its functionality.

CRITICAL REQUIREMENTS:
1. The function signature MUST remain EXACTLY the same: {signature}
2. The function MUST produce identical outputs for all possible inputs
3. The function MUST have identical side effects (exceptions, state changes, etc.)
4. The rewritten code MUST be syntactically valid Java code

FOCUS ON: Maximum transformation across all dimensions
- Apply control flow restructuring (loops, conditionals, branches)
- Rename variables and reorganize computations
- Transform the algorithmic approach while preserving semantics
- Use different Java language features and APIs
- Restructure the entire function logic flow
- Make the code as different as possible from the original

IMPORTANT: Do NOT add unnecessary code just for the sake of change. Every transformation must:
- Serve a legitimate structural purpose
- Maintain readability
- Preserve exact functionality
\end{lstlisting}

\begingroup
\makeatletter
\let\oldalgofinish\@algocf@finish
\renewcommand{\@algocf@finish}{%
    \oldalgofinish
    \par\vspace*{5cm}%
}
\makeatother

\begin{algorithm}[!h]
\small
\caption{FCGA-S: Subgraph-Level Refactoring}
\label{alg:refactor}
\KwIn{
    Graph $G=($ \\
    \Indp
    \Indp
    vertices $V=\{v_i\}$, \\
    edges $E=\{(v_i, v_j)\}$, \\
    \Indm
    $), G\in\mathcal{G}$, \\
    source codes $S=\{s_i\}, s_i\in\mathcal{S}$, \\
    node selection $R=\{r_k\}, R\in\mathcal{R}$, \\
    summarization LLM $K:\mathcal{S}\rightarrow\tilde{\mathcal{S}}$, \\
    refactoring LLM $L:\mathcal{G}\times\tilde{\mathcal{S}}^m\times\mathcal{R}\rightarrow\mathcal{G}\times\tilde{\mathcal{S}}^n$, \\
    embedding model $M:(\mathcal{S}\cup\tilde{\mathcal{S}})\rightarrow\mathbb{R}^d$
}
\KwOut{Augmented graph $G'$}
\tcp{Refactoring node set}
$V_R\leftarrow \{v_{r_k}\;|\;r_k\in R\}$\;
\tcp{And their summaries}
$\tilde{S}_R\leftarrow \{K(s_{r_k})\;|\;r_k\in R\}$\;
\tcp{Context node set}
$V_R^{(1)}\leftarrow \{v_j\;|\; \exists v_i\in V_R: (v_i, v_j)\in E\}$\;
\tcp{And their summaries}
$\tilde{S}_R^{(1)}\leftarrow \{K(s_{j})\;|\; \exists v_i\in V_R: (v_i, v_j)\in E\}$\;
\tcp{Build inducted subgraph}
$E_R\leftarrow \{(v_i, v_j)\;|\; (v_i, v_j)\in E, v_i\in V_R, v_j\in V_R \}$\;
\tcp{Context edge set}
$E_R^{(1)}\leftarrow \{(v_i, v_k)\,|\, (v_i, v_k)\in E, v_i\in V_R, v_k\in V_R^{(1)}\}$\;
\While{\textup{true}} {
\tcp{Refactor}
$((V_*, E_*), \tilde{S}_*)\leftarrow L((V_R\cup V_R^{(1)}, E_R\cup E_R^{(1)}), \tilde{S}_R\cup\tilde{S}_R^{(1)}, R)$\;
\tcp{Ensure context is preserved}
\uIf{
    $V_R^{(1)}\subset V_*$ \textbf{\textup{and}} $\exists\tilde{s}_{l}\in\tilde{S}_*\;\forall v_{l}\in V_*\setminus V_R^{(1)}$
}{\textbf{break}\;}
}
\tcp{Build the augmented graph}
$V'\leftarrow V\setminus V_R \cup V_*$\;
\tcp{Filter invalid edges}
$E'_*\leftarrow E_*\setminus \{(v_i, v_j)\;|\;v_i\in V_R^{(1)}, v_j\in V_R^{(1)}\}$\;
$E'\leftarrow E\setminus E_R\setminus E_R^{(1)} \cup E'_*$\;
\ForEach{$v_i$ in V'}{
    \uIf{$v_i\in V$}{
        \tcp{Guaranteed to exist in $S$}
        $X'[i]\leftarrow M(s_i)$\;
    }
    \Else{
        \tcp{Guaranteed to exist in $S_*$}
        $X'[i]\leftarrow M(\tilde{s}_i)$\;
    }
}
\Return{$G'=(V',E',X')$}\;
\end{algorithm}
\endgroup 

\newpage

\subsubsection{Subgraph-Level Refactoring}
\begin{lstlisting}[style=promptstyle]
You are a code refactoring engine specializing in Android applications. Your task is to accept a partial Function Call Graph (subgraph) and refactor specific target nodes while STRICTLY preserving the integrity of their adjacent nodes.

INSTRUCTIONS:
1. **Refactor**: Rewrite the logic of the whole set of nodes you are asked to refactor. YOU HAVE CREATIVE FREEDOM here.
  - **Example Strategies**:
     * **Splitting**: Break a complex function into multiple smaller helper nodes.
     * **Merging**: Combine a simple group of functions into a single node.
     * **Obfuscation**: Add dummy control flow (e.g., new functions that do meaningful nothing), or rename operations.
     * **Logic Hiding**: Move sensitive string construction into a new dedicated helper node.
     * **Other**: Any other strategy that you think is appropriate.
  - You can combine strategies, or even rewrite the entire set of functions.
2. **Preserve Functionalities**: All nodes that are not in the refactoring set MUST keep their original functionality and signature even if they are changed to accommodate the refactoring. They MUST be present in the final output graph.
3. **Reconstruct**: Generate the correct JSON of the new refactored program.

CONSTRAINTS & FORMATTING:
- **Output Format**: ONLY provide the final refactored graph in JSON. Do NOT include any other text, introduction, thinking, or explanation.
- **JSON Schema**: {nodes: [{id: str, signature: str, summary: str}], edges: [[source, target]]}.
  - **Node IDs**: Every node in the output graph MUST have an `id`.
    - For refactored or new nodes: Give a new unique ID starting with the letter R.
    - For context nodes: Copy the original node ID verbatim.
  - **Signatures**: Every node in the output graph MUST have a `signature`.
    - Function signatures MUST match exactly for non-refactored nodes.
  - **Summary**: Every node in the output graph MUST have a `summary`.
    - For updated nodes: Write a new summary of the new implementation.
    - For unchanged nodes: Copy the original summary verbatim.
- **No Code**: Do not output Java/Kotlin source code.
- **No Curly Braces**: Do NOT use curly braces `{}` in your summaries to simplify parsing.
\end{lstlisting}

\end{document}